\documentclass[twocolumn,aps,showpacs,superscriptaddress]{revtex4-1}
\usepackage{graphicx}
\usepackage{dcolumn}
\usepackage{bm}
\usepackage{color}
\begin{document}
\def\Journal#1#2#3#4{{#1} {\bf #2}, #3 (#4)}

\def\NCA{Nuovo Cimento}
\def\NIM{Nucl. Instr. Meth.}
\def\NIMA{{Nucl. Instr. Meth.} A}
\def\NPB{{Nucl. Phys.} B}
\def\NPA{{Nucl. Phys.} A}
\def\PLB{{Phys. Lett.}  B}
\def\PRL{Phys. Rev. Lett.}
\def\PRC{{Phys. Rev.} C}
\def\PRD{{Phys. Rev.} D}
\def\ZPC{{Z. Phys.} C}
\def\JPG{{J. Phys.} G}
\def\CPC{Comput. Phys. Commun.}
\def\EPJ{{Eur. Phys. J.} C}
\def\PR{Phys. Rept.}
\def\PRV{Phys. Rev.}
\def\JHEP{JHEP}

\preprint{}
\title{Dielectron Azimuthal Anisotropy at mid-rapidity in Au+Au collisions at $\sqrt{s_{_{NN}}} = 200$
GeV}
\affiliation{AGH University of Science and Technology, Cracow, Poland}
\affiliation{Argonne National Laboratory, Argonne, Illinois 60439, USA}
\affiliation{University of Birmingham, Birmingham, United Kingdom}
\affiliation{Brookhaven National Laboratory, Upton, New York 11973, USA}
\affiliation{University of California, Berkeley, California 94720, USA}
\affiliation{University of California, Davis, California 95616, USA}
\affiliation{University of California, Los Angeles, California 90095, USA}
\affiliation{Universidade Estadual de Campinas, Sao Paulo, Brazil}
\affiliation{Central China Normal University (HZNU), Wuhan 430079, China}
\affiliation{University of Illinois at Chicago, Chicago, Illinois 60607, USA}
\affiliation{Cracow University of Technology, Cracow, Poland}
\affiliation{Creighton University, Omaha, Nebraska 68178, USA}
\affiliation{Czech Technical University in Prague, FNSPE, Prague, 115 19, Czech Republic}
\affiliation{Nuclear Physics Institute AS CR, 250 68 \v{R}e\v{z}/Prague, Czech Republic}
\affiliation{Frankfurt Institute for Advanced Studies FIAS, Germany}
\affiliation{Institute of Physics, Bhubaneswar 751005, India}
\affiliation{Indian Institute of Technology, Mumbai, India}
\affiliation{Indiana University, Bloomington, Indiana 47408, USA}
\affiliation{Alikhanov Institute for Theoretical and Experimental Physics, Moscow, Russia}
\affiliation{University of Jammu, Jammu 180001, India}
\affiliation{Joint Institute for Nuclear Research, Dubna, 141 980, Russia}
\affiliation{Kent State University, Kent, Ohio 44242, USA}
\affiliation{University of Kentucky, Lexington, Kentucky, 40506-0055, USA}
\affiliation{Korea Institute of Science and Technology Information, Daejeon, Korea}
\affiliation{Institute of Modern Physics, Lanzhou, China}
\affiliation{Lawrence Berkeley National Laboratory, Berkeley, California 94720, USA}
\affiliation{Massachusetts Institute of Technology, Cambridge, Massachusetts 02139-4307, USA}
\affiliation{Max-Planck-Institut f\"ur Physik, Munich, Germany}
\affiliation{Michigan State University, East Lansing, Michigan 48824, USA}
\affiliation{Moscow Engineering Physics Institute, Moscow Russia}
\affiliation{National Institute of Science Education and Research, Bhubaneswar 751005, India}
\affiliation{Ohio State University, Columbus, Ohio 43210, USA}
\affiliation{Old Dominion University, Norfolk, Virginia 23529, USA}
\affiliation{Institute of Nuclear Physics PAN, Cracow, Poland}
\affiliation{Panjab University, Chandigarh 160014, India}
\affiliation{Pennsylvania State University, University Park, Pennsylvania 16802, USA}
\affiliation{Institute of High Energy Physics, Protvino, Russia}
\affiliation{Purdue University, West Lafayette, Indiana 47907, USA}
\affiliation{Pusan National University, Pusan, Republic of Korea}
\affiliation{University of Rajasthan, Jaipur 302004, India}
\affiliation{Rice University, Houston, Texas 77251, USA}
\affiliation{University of Science and Technology of China, Hefei 230026, China}
\affiliation{Shandong University, Jinan, Shandong 250100, China}
\affiliation{Shanghai Institute of Applied Physics, Shanghai 201800, China}
\affiliation{SUBATECH, Nantes, France}
\affiliation{Temple University, Philadelphia, Pennsylvania 19122, USA}
\affiliation{Texas A\&M University, College Station, Texas 77843, USA}
\affiliation{University of Texas, Austin, Texas 78712, USA}
\affiliation{University of Houston, Houston, Texas 77204, USA}
\affiliation{Tsinghua University, Beijing 100084, China}
\affiliation{United States Naval Academy, Annapolis, Maryland, 21402, USA}
\affiliation{Valparaiso University, Valparaiso, Indiana 46383, USA}
\affiliation{Variable Energy Cyclotron Centre, Kolkata 700064, India}
\affiliation{Warsaw University of Technology, Warsaw, Poland}
\affiliation{University of Washington, Seattle, Washington 98195, USA}
\affiliation{Wayne State University, Detroit, Michigan 48201, USA}
\affiliation{Yale University, New Haven, Connecticut 06520, USA}
\affiliation{University of Zagreb, Zagreb, HR-10002, Croatia}

\author{L.~Adamczyk}\affiliation{AGH University of Science and Technology, Cracow, Poland}
\author{J.~K.~Adkins}\affiliation{University of Kentucky, Lexington, Kentucky, 40506-0055, USA}
\author{G.~Agakishiev}\affiliation{Joint Institute for Nuclear Research, Dubna, 141 980, Russia}
\author{M.~M.~Aggarwal}\affiliation{Panjab University, Chandigarh 160014, India}
\author{Z.~Ahammed}\affiliation{Variable Energy Cyclotron Centre, Kolkata 700064, India}
\author{I.~Alekseev}\affiliation{Alikhanov Institute for Theoretical and Experimental Physics, Moscow, Russia}
\author{J.~Alford}\affiliation{Kent State University, Kent, Ohio 44242, USA}
\author{C.~D.~Anson}\affiliation{Ohio State University, Columbus, Ohio 43210, USA}
\author{A.~Aparin}\affiliation{Joint Institute for Nuclear Research, Dubna, 141 980, Russia}
\author{D.~Arkhipkin}\affiliation{Brookhaven National Laboratory, Upton, New York 11973, USA}
% ...

\collaboration{STAR Collaboration}\noaffiliation
\date{\today}
\begin{abstract}
We report on the first measurement of the azimuthal anisotropy
($v_2$) of dielectrons ($e^{+}e^{-}$ pairs) at mid-rapidity from $\sqrt{s_{_{NN}}} = 200$ GeV Au+Au
collisions with the STAR detector at RHIC, 
presented as a function of transverse momentum ($p_T$) for different invariant-mass
regions. In the mass region $M_{ee}\!<1.1$ GeV/$c^2$ the dielectron $v_2$ measurements are found to be consistent with expectations from $\pi^{0}$, $\eta$, $\omega$ and $\phi$ decay contributions.  In the mass region $1.1\!<M_{ee}\!<2.9$ GeV/$c^2$, the measured dielectron $v_2$ is consistent, within experimental  uncertainties, with that from the $c\bar{c}$ contributions.

\end{abstract}
\maketitle

\section{Introduction}
Dileptons are among the most essential tools for investigating the strongly interacting matter 
created in ultra-relativistic heavy-ion
collisions~\cite{starwhitepaper,otherwhitepapers}. Once produced, leptons, like photons, are not affected by the strong
interaction. Unlike photons, however, dileptons have an additional kinematic dimension: their invariant mass. Different kinematics of lepton pairs
[mass and transverse momentum ($p_T$) ranges] can selectively probe the
properties of the created matter throughout the whole
evolution~\cite{dilepton,dileptonII}.

In the low invariant mass range of produced lepton pairs
($M_{ll}\!<\!1.1$ GeV/$c^{2}$), vector mesons such as $\rho(770), 
 \omega(782)$, and $\phi(1020)$ and Dalitz decays of pseudoscalar mesons ($\pi^{0}$ and $\eta$) dominate the spectrum. In-medium properties of
the spectral functions of these vector mesons may exhibit modifications related to possible chiral
symmetry restoration~\cite{dilepton,dileptonII}, which can be studied via their dilepton decays. At SPS, the low-mass
dilepton enhancement in the CERES $e^+e^-$ data
~\cite{ceres} and in the NA60 $\mu^+\mu^-$ data~\cite{na60} could be attributed to  
substantial medium modification of the $\rho$-meson spectral
function. Two different realizations of chiral symmetry restoration were proposed: a dropping-mass scenario~\cite{dropmass}  and a broadening of the $\rho$ spectral function~\cite{massbroaden}, both of which described the CERES data. The precise NA60 measurement has
provided a decisive discrimination between the two scenarios, with only the broadened spectral function~\cite{massbroadenII} being able to describe the data.

At RHIC, a significant enhancement in the dielectron continuum,
compared to expectations from hadronic sources for
$0.15\!<M_{ee}\!<\!0.75$ GeV/$c^{2}$, was observed by the PHENIX Collaboration in
Au+Au collisions at $\sqrt{s_{_{NN}}} = 200$ GeV~\cite{lowmass}.
This enhancement is reported to increase from peripheral to central Au+Au
collisions and has a strong $p_T$ dependence. At low $p_T$ (below 1 GeV/$c$), the enhancement factor increases
from $1.5\pm0.3^{\textrm{stat}}\pm0.5^{\textrm{syst}}\pm0.3^\textrm{{model}}$ in 60-92\% peripheral Au+Au collisions to $7.6\pm0.5^\textrm{{stat}}\pm1.3^\textrm{{syst}}\pm1.5^\textrm{{model}}$ in 0-10\% central Au+Au
collisions. The last error is an estimate of the uncertainty in 
the extracted yield due to known hadronic sources. The STAR Collaboration
recently reported dielectron spectra in Au+Au collisions at $\sqrt{s_{_{NN}}} = 200$ GeV,
demonstrating an enhancement with respect to the contributions from known hadronic sources in the low mass region that bears little centrality dependence~\cite{staromega}. Theoretical calculations~\cite{rapp:09,PSHD:12,USTC:12}, which describe the SPS dilepton data, 
fail to consistently describe the low-$p_T$ and low-mass
enhancement observed by PHENIX in 0-10\% and 10-20\% central Au+Au collisions~\cite{lowmass}. The same calculations, however, describe the STAR measurement of the low-$p_T$ and low-mass enhancement from peripheral to central Au+Au collisions~\cite{staromega}. 

For $1\!<\!p_{T}\!<\!5$ GeV/$c$ and in the mass region $M_{ee}\!<\!0.3$ GeV/$c^{2}$, the PHENIX Collaboration
derived direct photon yields through dielectron measurements to assess thermal radiation at RHIC~\cite{thermalphoton}.
The excess of direct photon yield in central Au+Au collisions over that observed in $p+p$ collisions is found to fall off exponentially with $p_T$ with an inverse slope of 220 MeV/$c$. In addition, the azimuthal anisotropy $v_2$, the second
harmonic of the azimuthal distribution with respect to the
event plane~\cite{Art:98}, has been measured for direct photons using electro-magnetic calorimeter  and
found to be substantial and comparable to the $v_2$ for hadrons for $1\!<\!p_{T}\!<\!4$
GeV/$c$~\cite{photonv2}. Model calculations for thermal photons from the Quark-Gluon Plasma (QGP) in this kinematic region significantly under-predict the
observed $v_2$, while the model calculations which include a
significant contribution from the hadronic sources at a later stage
describe the excess of the spectra and the substantial $v_2$ for
$1\!<\!p_{T}\!<\!4$ GeV/$c$ reasonably well~\cite{rapp:11}.
With their augmented kinematics,
dilepton $v_2$ measurements have been proposed as
an alternative study of medium properties~\cite{Gale:07}.
Specifically, the $v_2$ as a function of $p_T$ in different invariant mass regions will enable us to probe the
properties of the medium at different stages, from QGP to hadron-gas dominated.

The dilepton spectra in the intermediate mass range
($1.1\!<M_{ll}\!<\!3.0$ GeV/$c^{2}$) are expected to be 
related to the QGP thermal radiation~\cite{dilepton,dileptonII}.
However, contributions from other sources have to be measured
experimentally, e.g. electron or muon pairs from semileptonic decays of open
charm or bottom hadrons ($c+\bar{c}\rightarrow l^{+}+l^{-}X$ or
$b+\bar{b}\rightarrow l^{+}+l^{-}X$). Utilizing dielectrons, the PHENIX Collaboration obtained the charm and bottom cross sections in p+p collisions at $\sqrt{s} = 200$ GeV~\cite{phenixcharm:09}.

 With the installation of a Time-of-Flight (TOF)
detector~\cite{startof}, as well as an upgrade of the data acquisition
system~\cite{stardaq}, the STAR detector with its large acceptance
provides excellent electron identification capability at low
momentum for dielectron analyses~\cite{star}. 

In this paper, we present the
first dielectron $v_2$ measurements from low to intermediate mass ($M_{ee}\!<\!2.9$
GeV/$c^{2}$) in Au+Au collisions at $\sqrt{s_{_{NN}}} = 200$ GeV.
This paper is organized as follows. Sect.~\ref{detector} describes
the detector and data samples used in the analysis.
Sects.~\ref{identification} and~\ref{bg} describe the electron
identification, electron pair distributions, and background
subtraction. Sects.~\ref{flowmethod} and~\ref{simu} describe the analysis
details of the azimuthal anisotropy and simulation. Sect.~\ref{sys} describes the systematic uncertainties.
Results for the centrality, mass, and $p_T$ dependence of dielectron $v_2$
are presented in detail in Sect.~\ref{results}. Lastly,
Sect.~\ref{summary} provides a concluding summary.

\section{Detector And Data Sample}\label{detector} The two main detectors
used in this analysis are the
Time Projection Chamber (TPC)~\cite{startpc} and the TOF
detector. Both have full azimuthal coverage
at mid-rapidity. The TPC is STAR's main tracking detector,
measuring momentum, charge, and energy loss of charged particles.
The ionization energy loss ($dE/dx$) of charged particles in the TPC gas
is used for particle identification~\cite{bichsel,pidpp08}. In addition, the TOF
detector extends STAR's hadron identification capabilities to higher momenta and significantly improves its electron identification capabilities~\cite{pidNIMA,tofPID}.

\renewcommand{\floatpagefraction}{0.75}
\begin{figure*}[htbp]
\begin{center}
\includegraphics[keepaspectratio,width=0.99\textwidth]{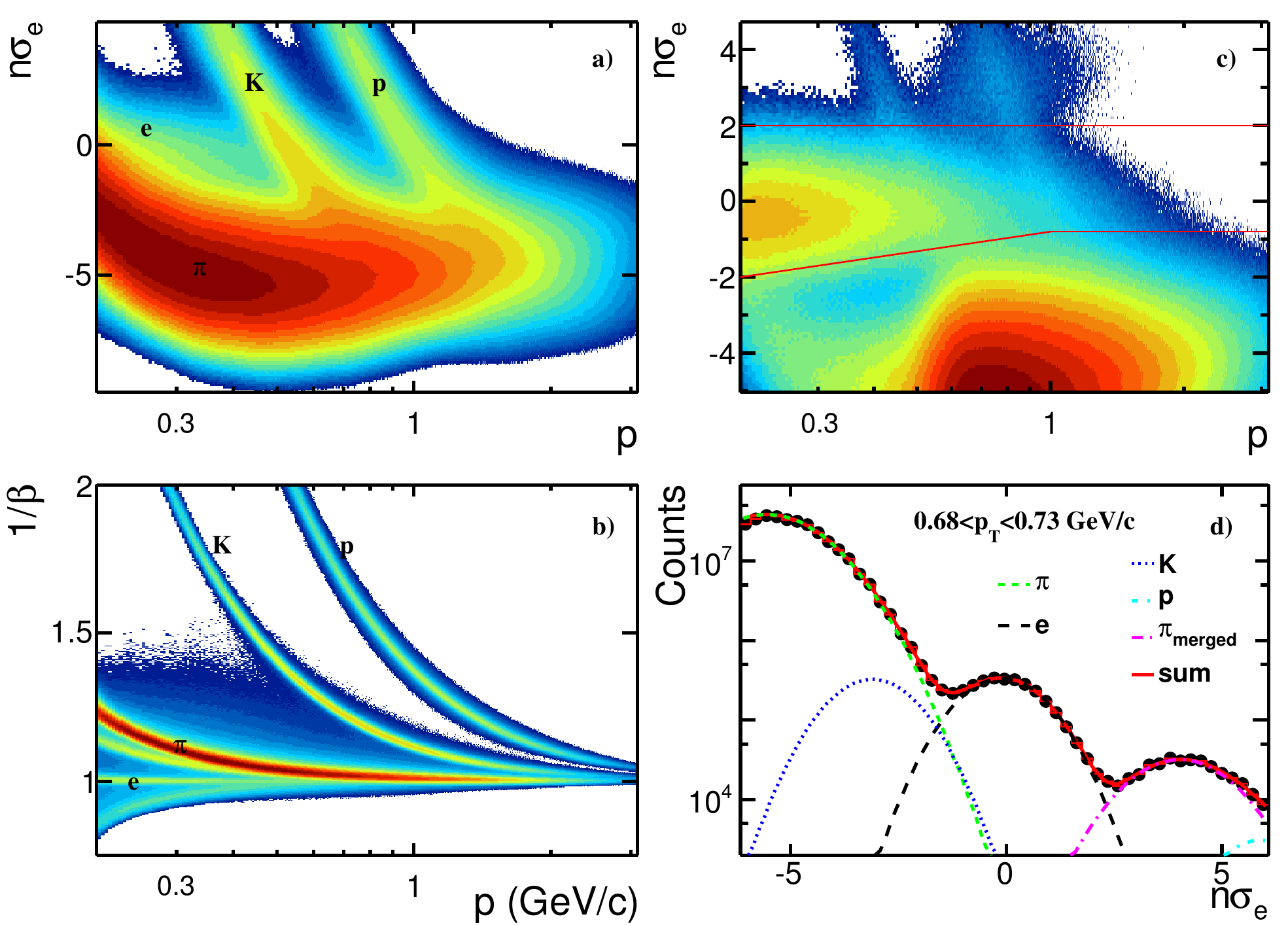}
\caption{(Color online) Panel (a): The normalized
$dE/dx$ distribution as a function of momentum from  TPC in Au+Au collisions at $\sqrt{s_{_{NN}}} = 200$ GeV. Panel (b): $1/\beta$ measurements from TOF versus the
momentum from  TPC in Au+Au collisions. The $1/\beta$ resolution is 0.011. Panel
(c): The normalized $dE/dx$ distribution as a
function of momentum with the cut of
$|1/\beta-1/\beta_{\textrm{exp}}|\!<\!0.025$. An electron band is prominent
with the requirement of velocity close to the speed of light from the
TOF measurement. Electron
candidates whose $n\sigma_{e}$ falls between the lines are selected for further dielectron analysis.  Panel (d):  The $n\sigma_{e}$ distribution for
$0.68\!<\!p_T\!<\!0.73$ GeV/$c$ after the cut 
$|1/\beta-1/\beta_{\textrm{exp}}|\!<\!0.025$ is applied. The solid curve represents a
multiple Gaussian fit to the $n\sigma_{e}$ distribution. Different components from the fit are also shown. The $\pi_{merged}$ represents contribution from two merged $\pi$ tracks. 
} \label{pid}
\end{center}
\end{figure*}

The data used for this analysis were taken in 2010 and 2011. A total of 760 million minimum-bias events, with 240 million from 2010 and 520 million from 2011 data samples of $\sqrt{s_{_{NN}}} = 200$ GeV Au+Au collisions were used in the analysis. These events were required to have
collision vertices within 30 cm of the TPC
center along the beam line, where the material budget is minimal (0.6\% in radiation length in front of the TPC inner field cage). The
minimum-bias trigger was defined by the coincidence of signals from
the two Vertex Position Detectors (VPDs)~\cite{starvpd}, located
on each side of the STAR barrel, covering a pseudorapidity range of $4.4<|\eta|<4.9$. The centrality tagging was determined by the measured charged particle multiplicity density in the TPC within $|\eta|<0.5$~\cite{centraltag}. The 2010 and 2011 minimum-bias data (0-80\% centrality) were analyzed separately. The dielectron $v_2$ measurement in this article is the combined $v_2$ result from these two data sets. 

\section{Data Analysis}\label{analysis}
\subsection{Electron identification}\label{identification}
\begin{figure}[htbp]
\begin{center}
\includegraphics[keepaspectratio,width=0.45\textwidth]{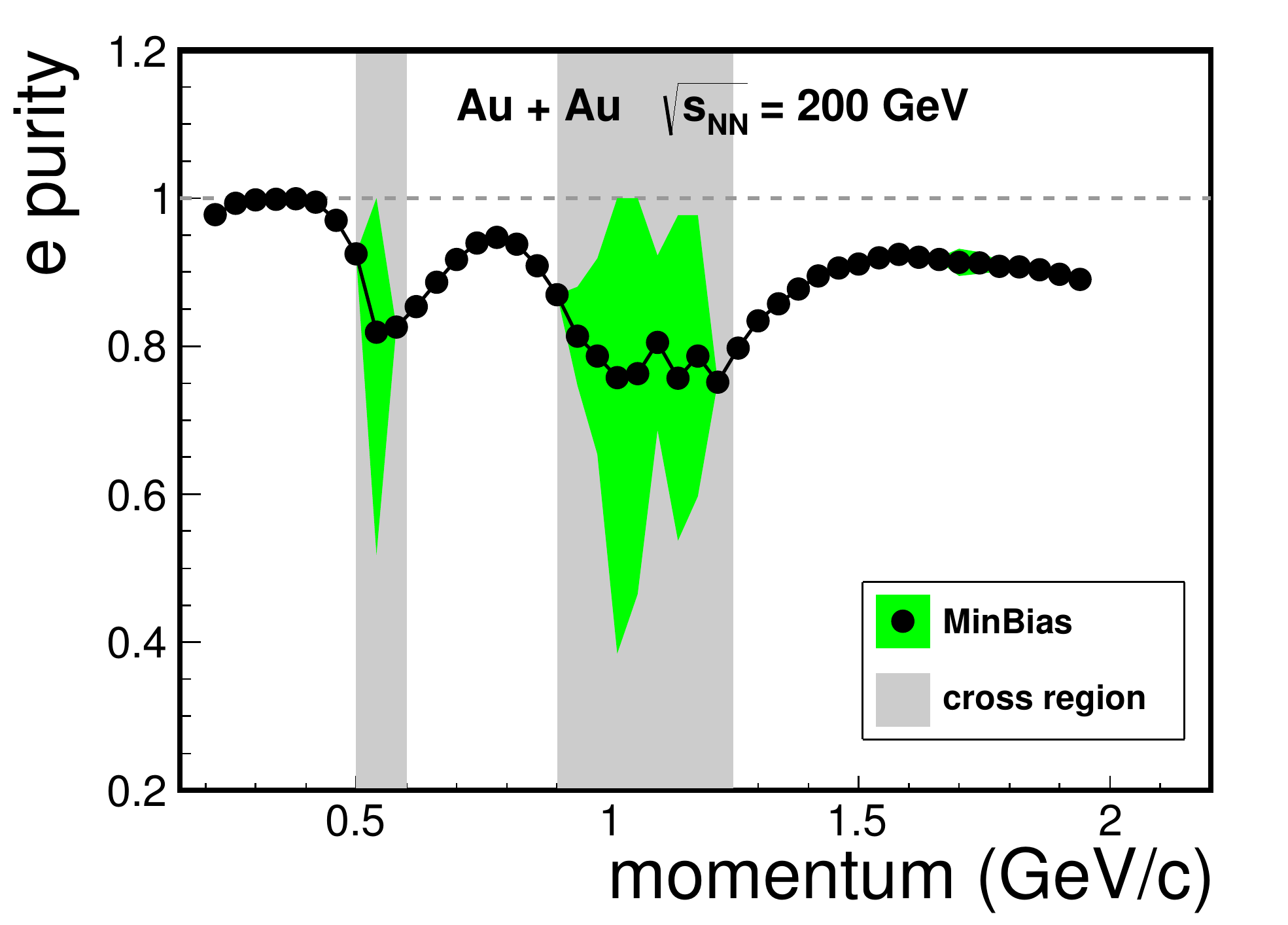}
\caption{(Color online) The purity of electron candidates as a function of momentum in minimum-bias Au+Au
collisions at $\sqrt{s_{_{NN}}} = 200$ GeV. In the cross regions, the electron candidates overlap with hadron components in the $dE/dx$ distribution, which results in large uncertainties in the multi-component fit, as shown by the shading around the data points. } \label{purity}
\end{center}
\end{figure}

Particles directly originating from the collision, with
trajectories that project back to within 1 cm of the collision
vertex, calculated in three dimensions, were selected for this analysis. Table~\ref{tab:I} lists
selection criteria for the tracks for further electron
identification.
\begin{table}\caption{Criteria used for the selection of tracks for electron identification. NFit is
the number of points used to fit the TPC track, and NMax is
the maximum possible number for that track. $dE/dx$ points is the
number of points used to derive the $dE/dx$ value. The DCA is the
distance of the closest approach between the trajectory of a particle
and the collision vertex. \label{tab:I}}
{\centering
\begin{tabular}{c|c} \hline\hline 
 $|\eta|$ & $<$ 1\\
$p_T$& $>$ 0.2 GeV/$c$\\
DCA & $<$ 1 cm\\
NFit & $>$ 19 \\
 NFit / NMax & $>$ 0.52 \\
$dE/dx$ points & $>$ 15 \\
\hline \hline
\end{tabular}
}
\end{table}
 The normalized
$dE/dx$ ($n\sigma_{e}$) is
defined as: $n\sigma_{e}=\ln(dE/dx/I_{e})/R_{e}$, where $dE/dx$ is
the measured energy loss of a particle, and $I_{e}$ is
the expected $dE/dx$ of an electron. $R_{e}$ is the resolution of $\ln(dE/dx/I_{e})$, defined as the width of its distribution, and is better than
8\% for these data. Figure~\ref{pid} panel (a) shows the $n\sigma_{e}$ distribution as a function of momentum from the TPC, while panel (b) shows the inverse velocity $1/\beta$ measurements from the TOF versus the
momentum measured by the TPC. Panel (c) shows the $n\sigma_{e}$ distribution versus momentum with the requirement on velocity that $|1/\beta-1/\beta_{\textrm{exp}}|\!<\!0.025$, in which $\beta_{\textrm{exp}}$ is the velocity calculated with the assumption of electron mass. Panel (d) presents the $n\sigma_{e}$ distribution for $0.68\!<\!p_T\!<\!0.73$ GeV/$c$ after the cut  $|1/\beta-1/\beta_{\textrm{exp}}|\!<\!0.025$ is applied.  With perfect calibration, the $n\sigma_{e}$ for single
electrons should follow a standard normal distribution. Electron
candidates whose $n\sigma_{e}$ falls between the lines in
Fig.~\ref{pid} panel (c) are selected. From the multiple-component fit to the $dE/dx$ distribution, an example of which is shown in Panel (d), one can obtain the purity of electron candidates. The purity is 95\% on average and depends on momentum~\cite{staromega}, as shown in Fig~\ref{purity}. 
\renewcommand{\floatpagefraction}{0.75}
With the combined information of velocity
($\beta$) from the TOF and $dE/dx$ from the TPC,
electrons can be clearly identified from low to intermediate $p_T$
($0.2\!<\!p_T\!<\!3$ GeV/$c$) for
$|\eta|\!<\!1$~\cite{starelectron,starppdilepton}. This is important for dielectron measurements from low to intermediate mass region.

\subsection{Dielectron invariant mass distribution and background subtraction}\label{bg}
\renewcommand{\floatpagefraction}{0.75}
\begin{figure}[htbp]
\begin{center}
\includegraphics[keepaspectratio,width=0.50\textwidth]{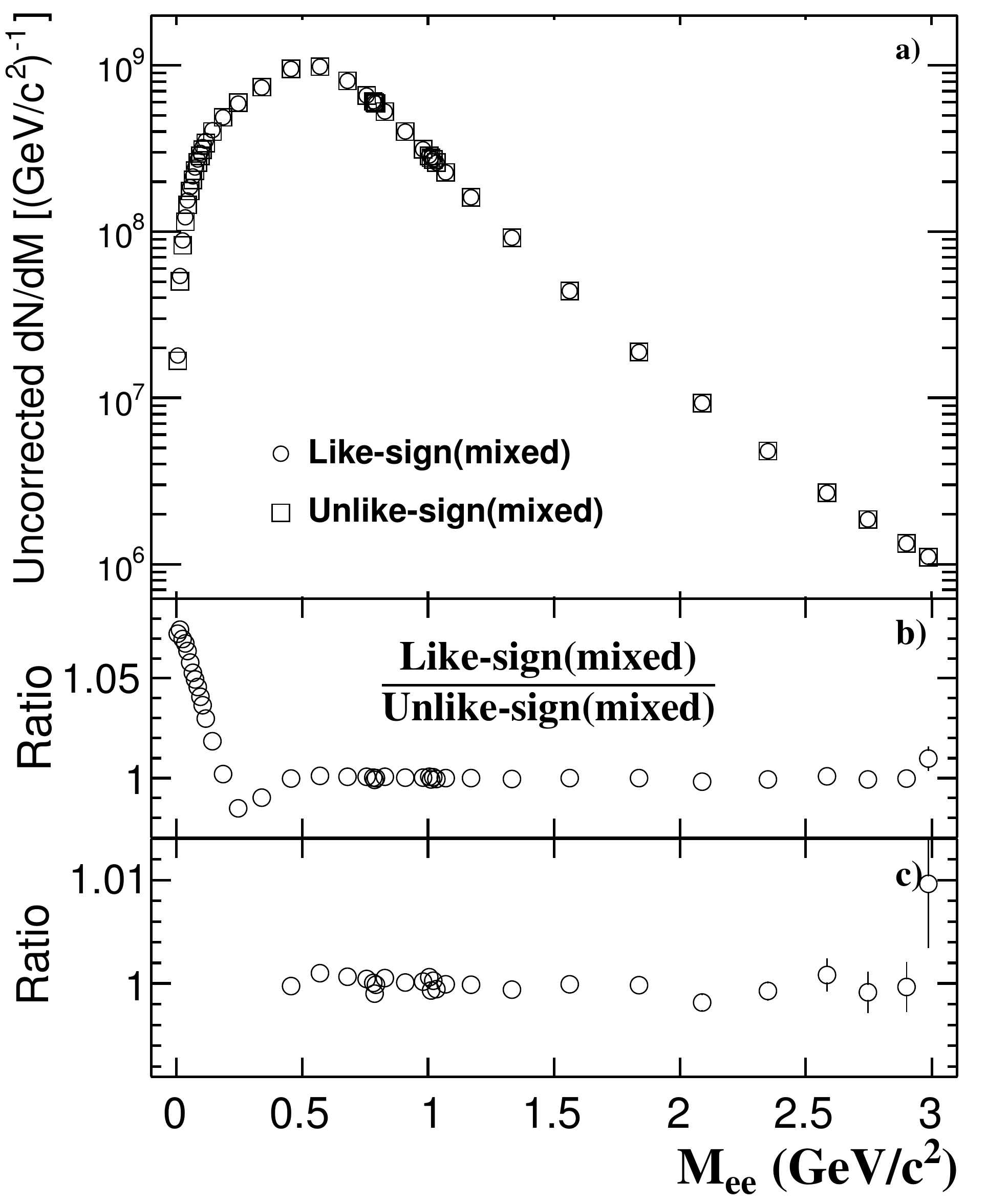}
\caption{(Color online) Panel (a): The mixed-event unlike-sign and
mixed-event like-sign electron pair invariant mass distributions in
minimum-bias Au+Au collisions at $\sqrt{s_{_{NN}}} = 200$ GeV.
Panel (b): The ratio of mixed-event like-sign distribution to
mixed-event unlike-sign distribution in minimum-bias Au+Au
collisions at $\sqrt{s_{_{NN}}} = 200$ GeV. Panel (c): A zoom-in version of Panel (b).} \label{accepplot}
\end{center}
\end{figure}

\renewcommand{\floatpagefraction}{0.75}
\begin{figure}[htbp]
\begin{center}
\includegraphics[keepaspectratio,width=0.50\textwidth]{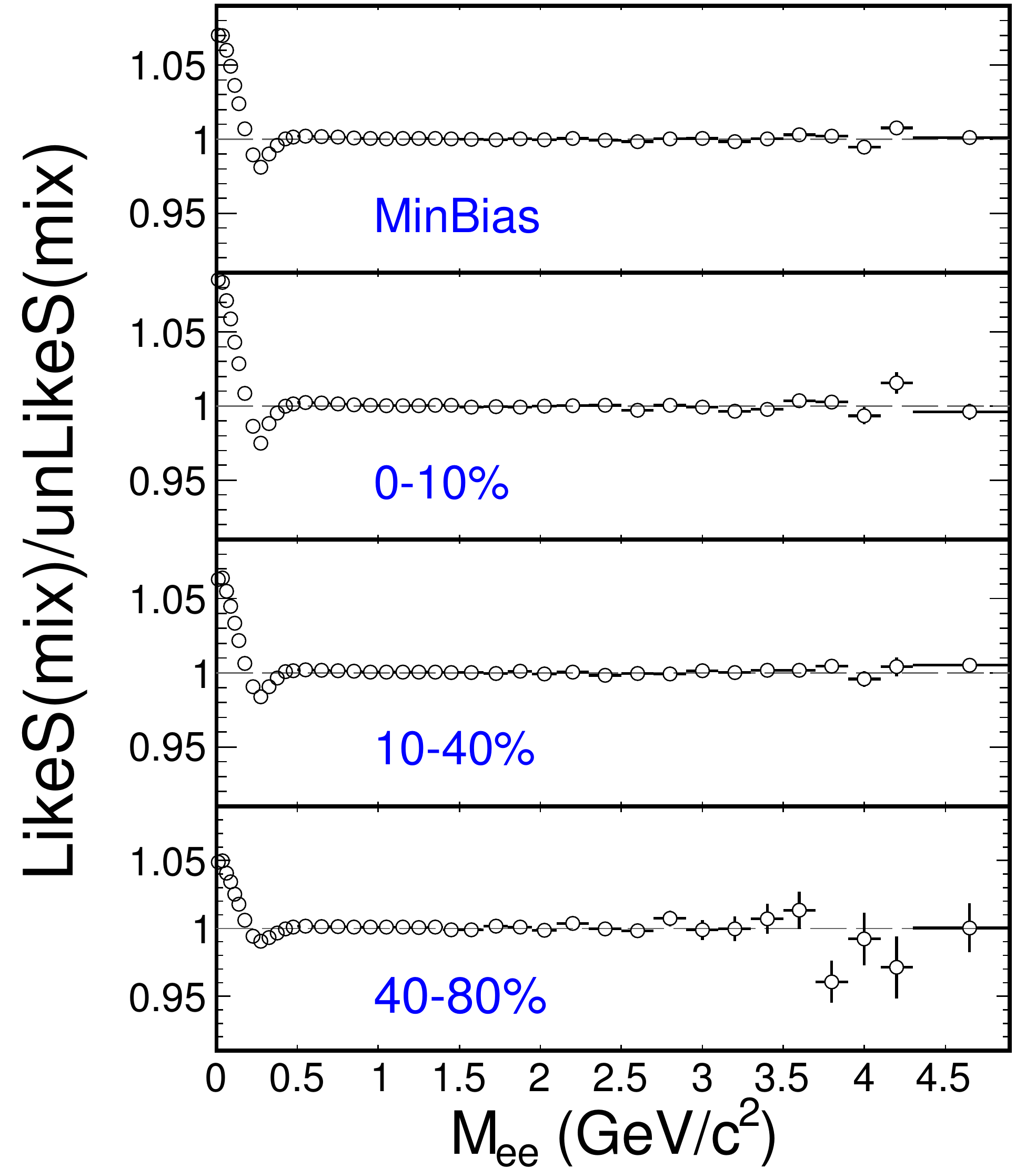}
\caption{(Color online) The ratio of the mixed-event like-sign distribution to the 
mixed-event unlike-sign distribution in minimum-bias, as well as specific centrality selections of, Au+Au collisions at $\sqrt{s_{_{NN}}} = 200$ GeV. } \label{accepratplot}
\end{center}
\end{figure}

\renewcommand{\floatpagefraction}{0.75}
\begin{figure}[htbp]
\begin{center}
\includegraphics[keepaspectratio,width=0.50\textwidth]{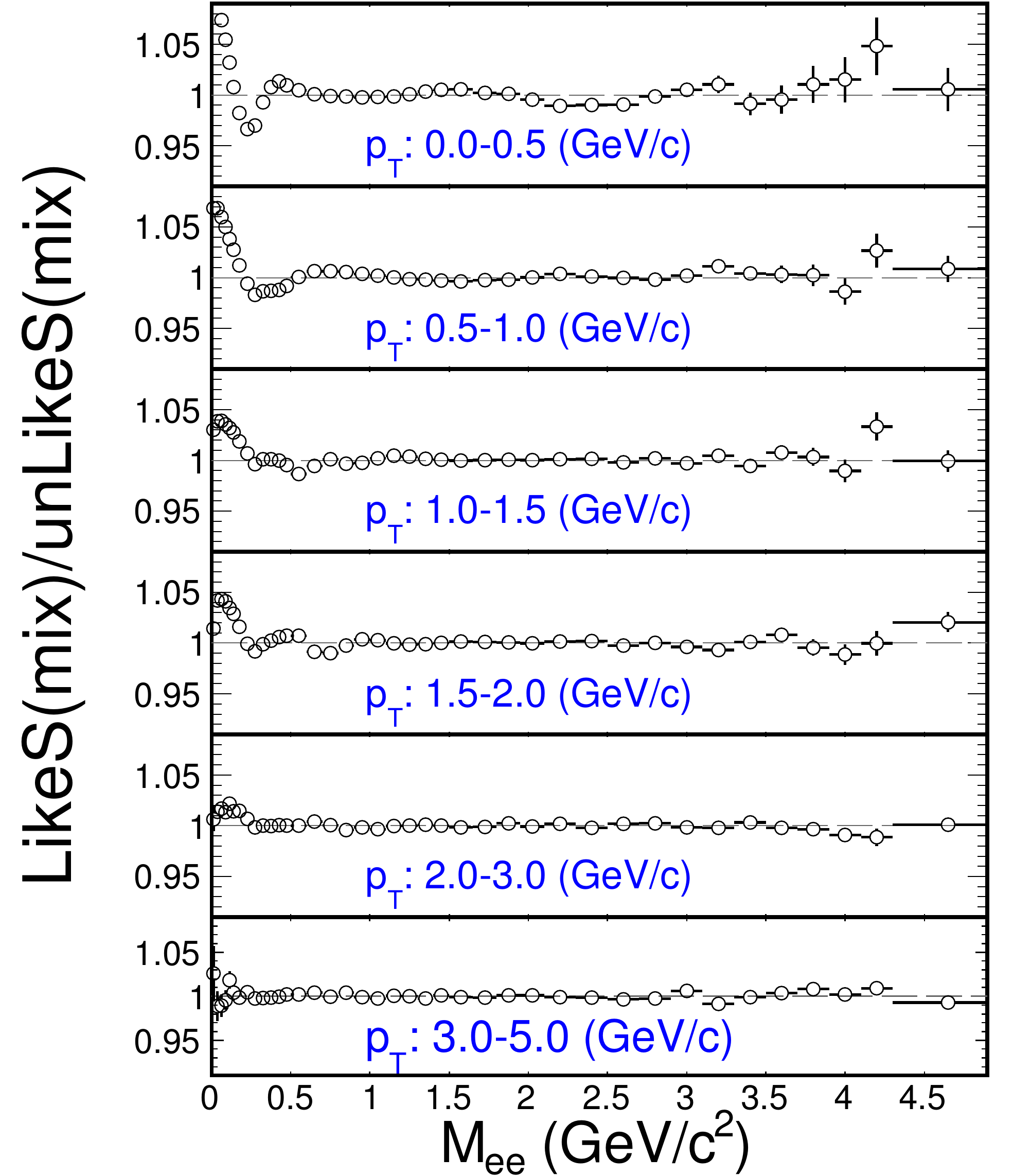}
\caption{(Color online) The ratio of the mixed-event like-sign distribution to the 
mixed-event unlike-sign distribution in different $p_T$ ranges in minimum-bias Au+Au
collisions at $\sqrt{s_{_{NN}}} = 200$ GeV. } \label{accepratptplot}
\end{center}
\end{figure}

The dielectron signals may come from decays of both light-flavor
and heavy-flavor hadrons. The light-flavor sources include $\pi^{0}$, $\eta$, and
$\eta^{\prime}$ Dalitz decays: $\pi^{0}\rightarrow \gamma
e^{+}e^{-}$, $\eta \rightarrow \gamma e^{+}e^{-}$, and
$\eta^{\prime}\rightarrow \gamma e^{+}e^{-}$; and vector meson decays:
$\omega \rightarrow \pi^{0} e^{+}e^{-}$, $\omega \rightarrow
e^{+}e^{-}$, $\rho^{0} \rightarrow e^{+}e^{-}$, $\phi \rightarrow
\eta e^{+}e^{-}$, and $\phi \rightarrow e^{+}e^{-}$. The heavy-flavor sources include $J/\psi
\rightarrow e^{+}e^{-}$ and heavy-flavor hadron semi-leptonic decays:
$c\bar{c} \rightarrow e^{+}e^{-}$ and $b\bar{b} \rightarrow
e^{+}e^{-}$. The signals also include Drell-Yan contributions. The dielectron contributions from 
photon conversions ($\gamma \rightarrow e^{+}e^{-}$) in the detector material
are present in the raw data. The momenta of these electrons are biased, which results in a multiple-peak structure in the dielectron mass distribution for $M_{ee}\!<\!0.12$ GeV/$c^2$. The peak position in the mass distribution depends on the conversion point in the detector~\cite{Jie:13}. It is found that the dielectron $v_2$ from photon conversions is the same as that from $\pi^{0}$ Dalitz decays. The vector meson contributions to the Au+Au data may be modified in the medium. QGP thermal radiation and additional contributions from the hadron gas would also be contained in the data.
\renewcommand{\floatpagefraction}{0.75}
\begin{figure}[htbp]
\begin{center}
\includegraphics[keepaspectratio,width=0.5\textwidth]{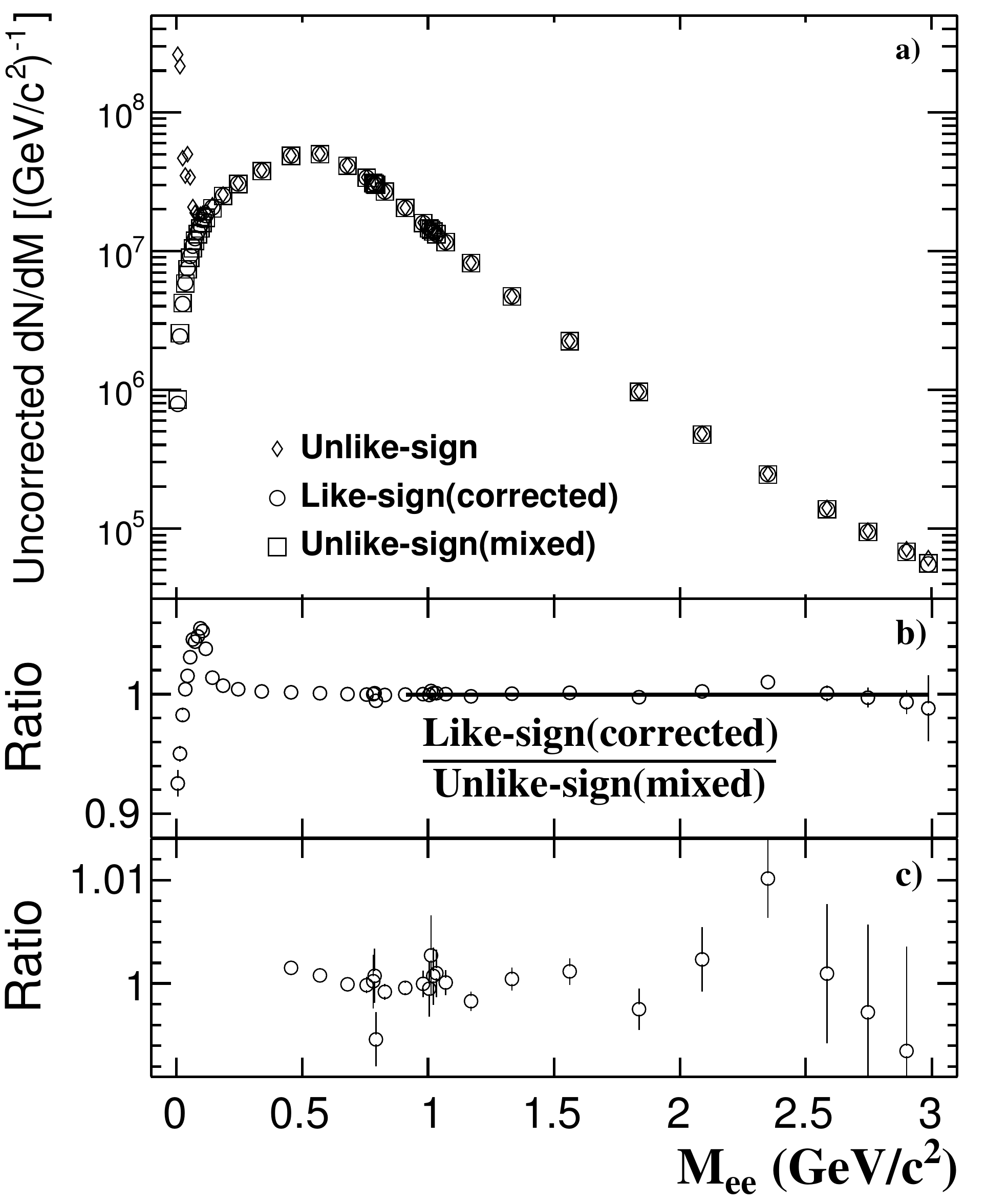}
\caption{(Color online) Panel (a): The electron pair invariant
mass distributions for same-event unlike-sign pairs, same-event
like-sign, and mixed-event unlike-sign in minimum-bias Au+Au
collisions at $\sqrt{s_{_{NN}}} = 200$ GeV. The electron
candidates are required to be in the range  $|\eta|\!<1$ and
have  $p_T$ greater than 0.2 GeV/$c$. The $ee$ pairs are
required to be in the rapidity range  $|y_{ee}|\!<\!1$.
Variable bin widths are used for the yields and  signal-to-background ratios. Panel (b): The ratio of the same-event like-sign distribution (corrected for the acceptance difference) to the normalized
mixed-event unlike-sign distribution in minimum-bias Au+Au
collisions at $\sqrt{s_{_{NN}}} = 200$ GeV. Panel (c): A zoom-in version of Panel (b).} \label{eepair}
\end{center}
\end{figure}

\renewcommand{\floatpagefraction}{0.75}
\begin{figure}[htbp]
\begin{center}
\includegraphics[keepaspectratio,width=0.5\textwidth]{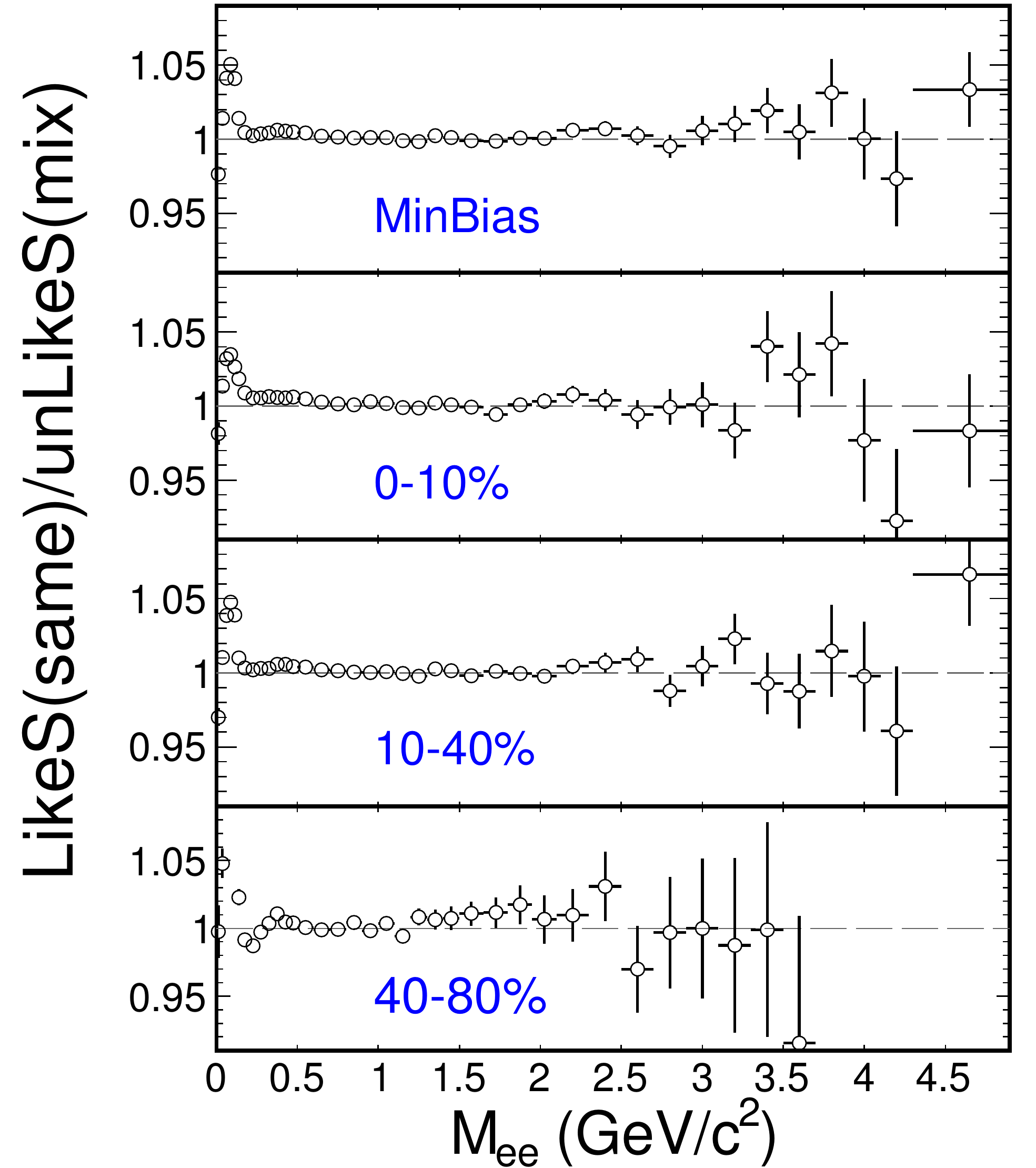}
\caption{(Color online) The centrality dependence of the ratio of the same-event like-sign distribution (corrected for the acceptance difference) to the normalized
mixed-event unlike-sign distribution in minimum-bias Au+Au
collisions at $\sqrt{s_{_{NN}}} = 200$ GeV. } \label{centrat}
\end{center}
\end{figure}

\renewcommand{\floatpagefraction}{0.75}
\begin{figure}[htbp]
\begin{center}
\includegraphics[keepaspectratio,width=0.5\textwidth]{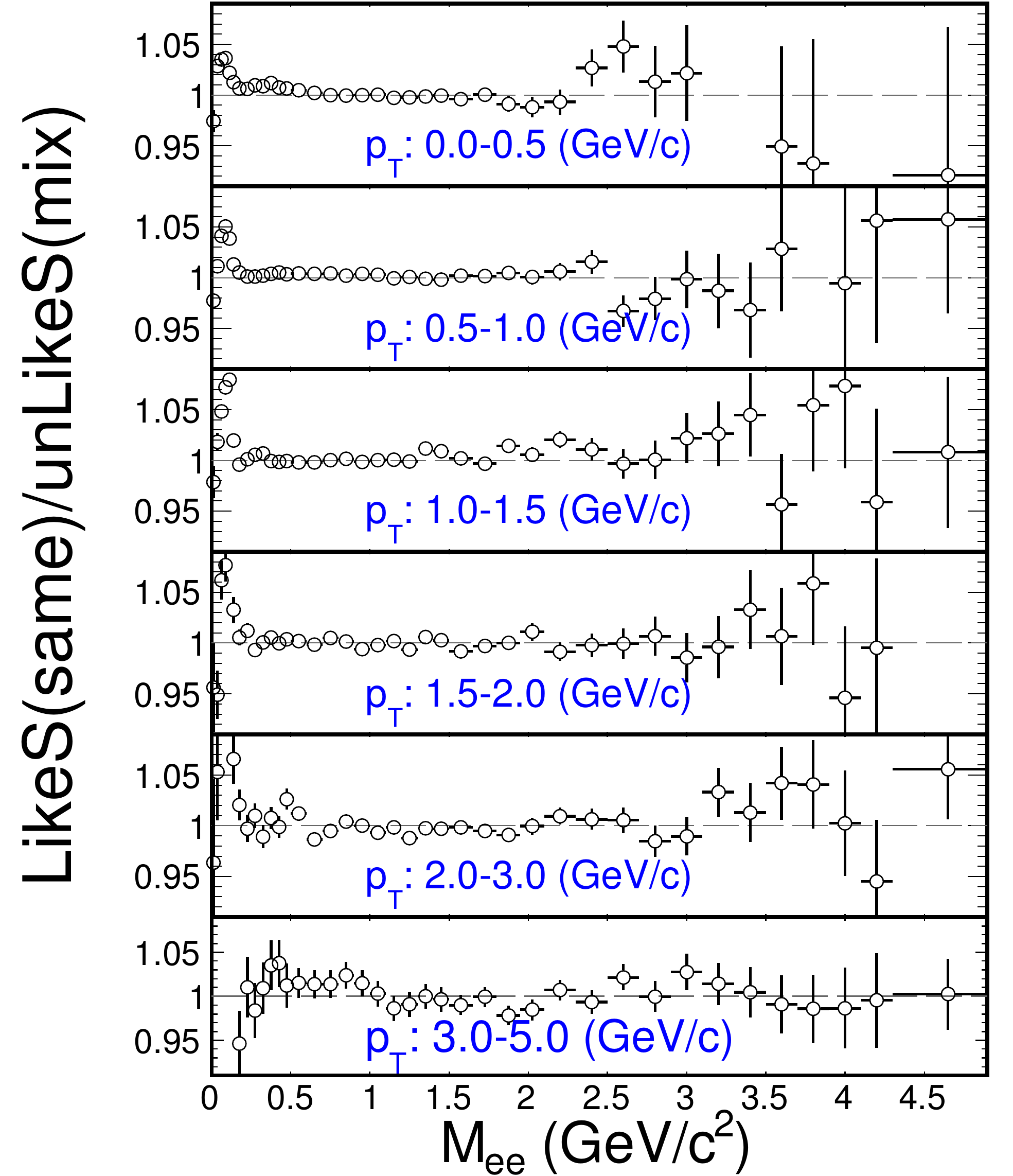}
\caption{(Color online) The $p_T$ dependence of the ratio of the same-event like-sign distribution (corrected for the acceptance difference) to the normalized
mixed-event unlike-sign distribution in minimum-bias Au+Au
collisions at $\sqrt{s_{_{NN}}} = 200$ GeV. } \label{ptrat}
\end{center}
\end{figure}

\renewcommand{\floatpagefraction}{0.75}
\begin{figure}[htbp]
\begin{center}
\includegraphics[keepaspectratio,width=0.5\textwidth]{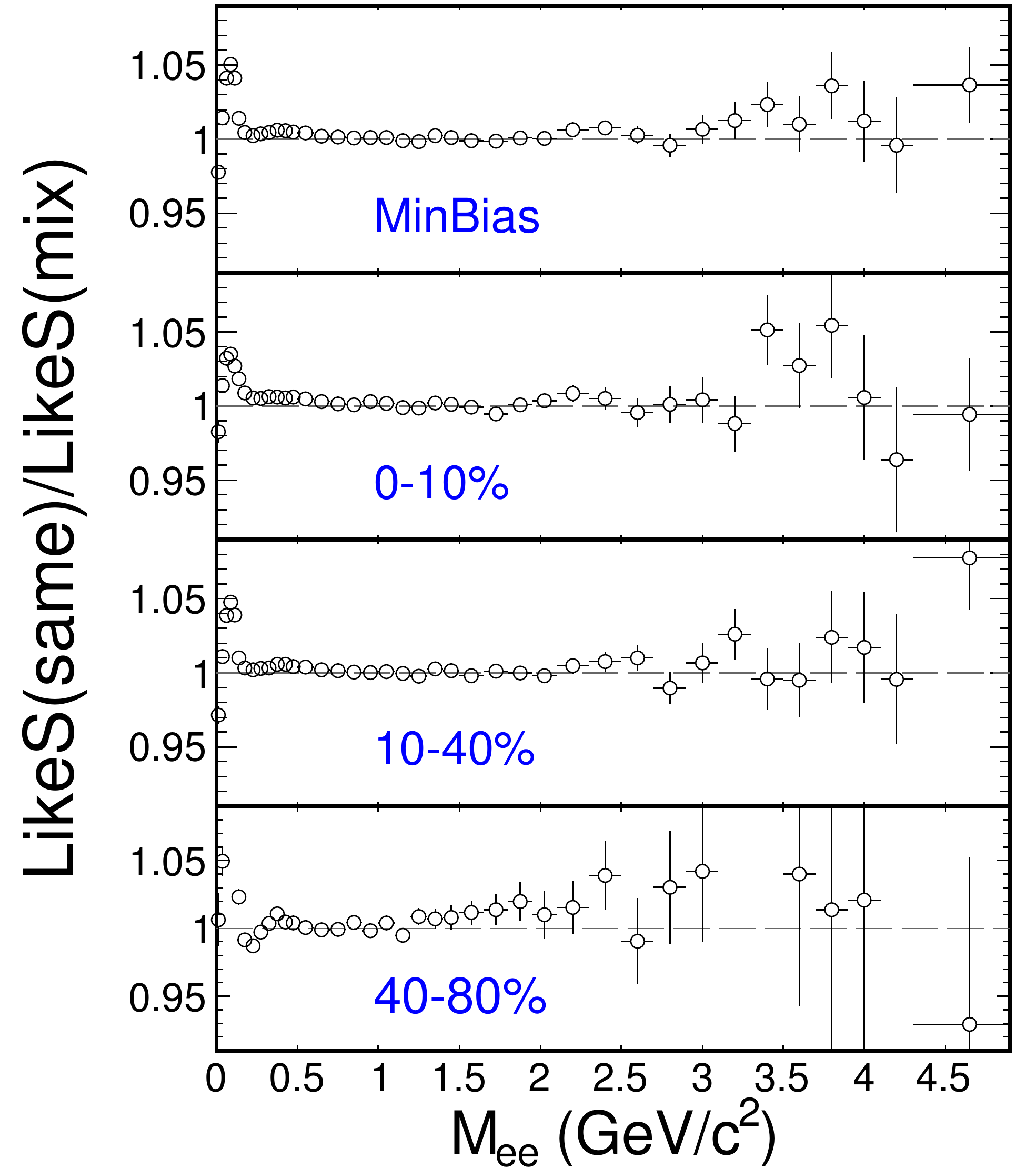}
\caption{(Color online) The centrality dependence of the ratio of the same-event like-sign distribution to the normalized mixed-event like-sign distribution in minimum-bias Au+Au
collisions at $\sqrt{s_{_{NN}}} = 200$ GeV. } \label{likesigncentrat}
\end{center}
\end{figure}

\renewcommand{\floatpagefraction}{0.75}
\begin{figure}[htbp]
\begin{center}
\includegraphics[keepaspectratio,width=0.5\textwidth]{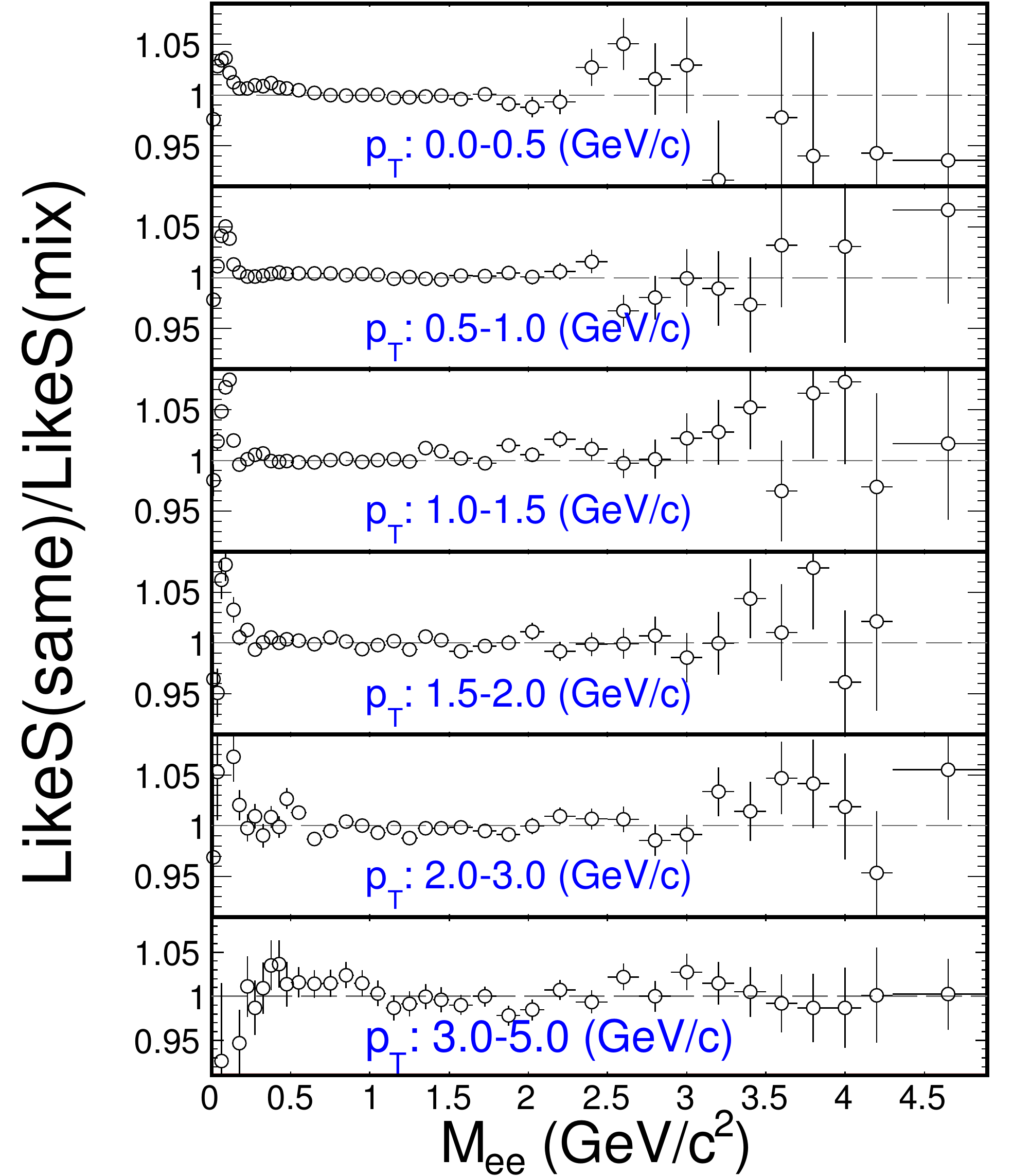}
\caption{(Color online) The $p_T$ dependence of the ratio of the same-event like-sign distribution to the normalized mixed-event like-sign distribution in minimum-bias Au+Au
collisions at $\sqrt{s_{_{NN}}} = 200$ GeV. } \label{likesignptrat}
\end{center}
\end{figure}

\renewcommand{\floatpagefraction}{0.75}
\begin{figure}[htbp]
\begin{center}
\includegraphics[keepaspectratio,width=0.5\textwidth]{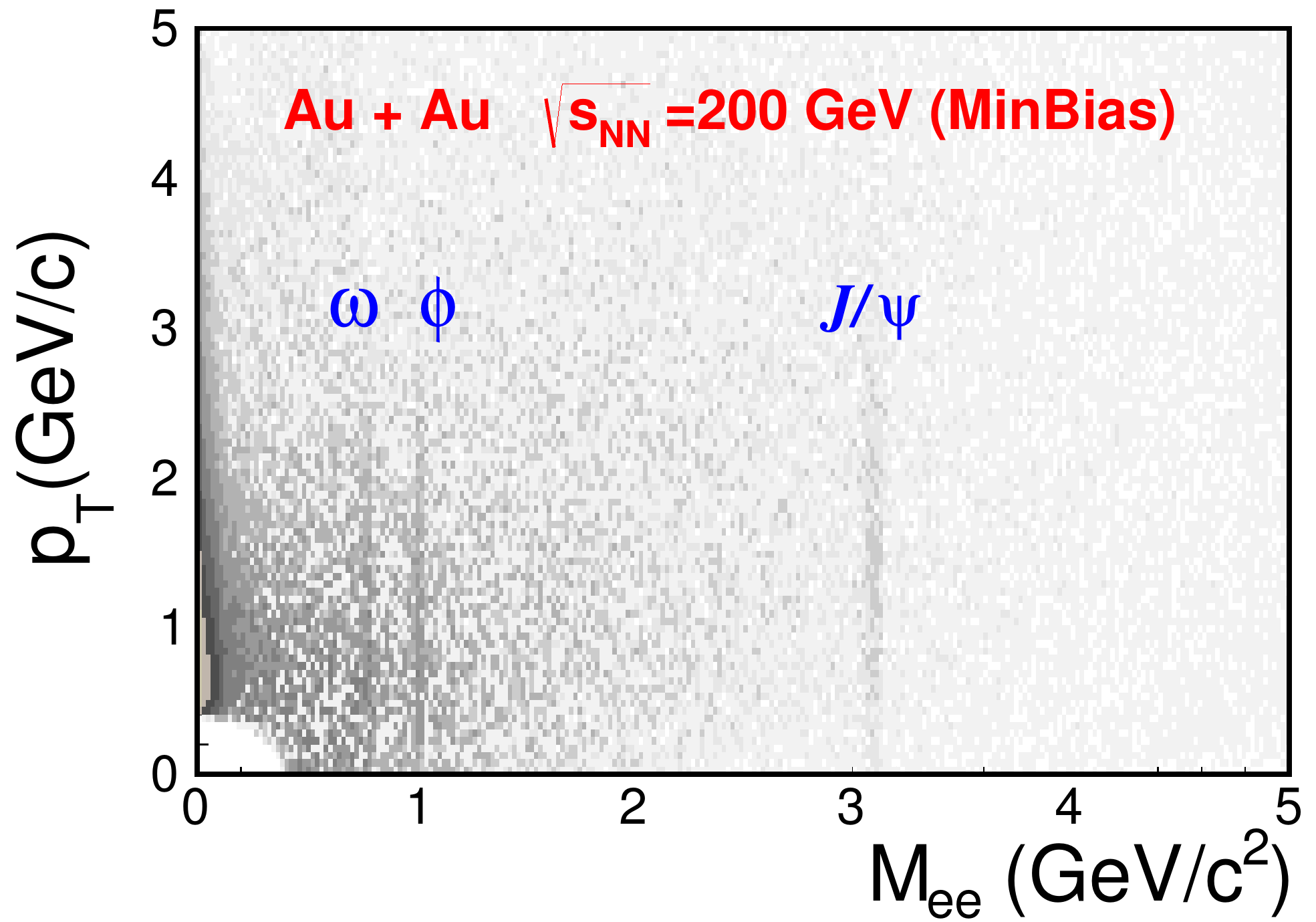}
\caption{(Color online) The $p_T$ as a function of $M_{ee}$ for dielectron signal
without efficiency correction in $\sqrt{s_{_{NN}}} = 200$ GeV
minimum-bias Au+Au collisions. }
\label{2drawcontinuum}
\end{center}
\end{figure}

With high purity electron samples, the $e^{+}e^{-}$
pairs from each event are accumulated to generate the invariant
mass distributions ($M_{ee}$), here referred to as the unlike-sign distributions.
The unlike-sign distributions contain both signal (defined in the previous paragraph)
and backgrounds of random combinatorial pairs and correlated cross pairs.
The correlated cross pairs come from two  $e^{+}e^{-}$ pairs from a single
meson decay: a Dalitz decay followed by a conversion of the decay
photon, or conversions of multiple photons from the same meson. The
electron candidates are required to be in the range 
$|\eta|\!<1$ and $p_T>0.2$ GeV/$c$, while the rapidity of $e^{+}e^{-}$ pairs ($y_{ee}$) is
required to be in the region $|y_{ee}|\!<\!1$.

Two methods are used for background estimation, based on
same-event like-sign and mixed-event unlike-sign techniques. 
\renewcommand{\floatpagefraction}{0.75}
\begin{figure}[htbp]
\begin{center}
\includegraphics[keepaspectratio,width=0.5\textwidth]{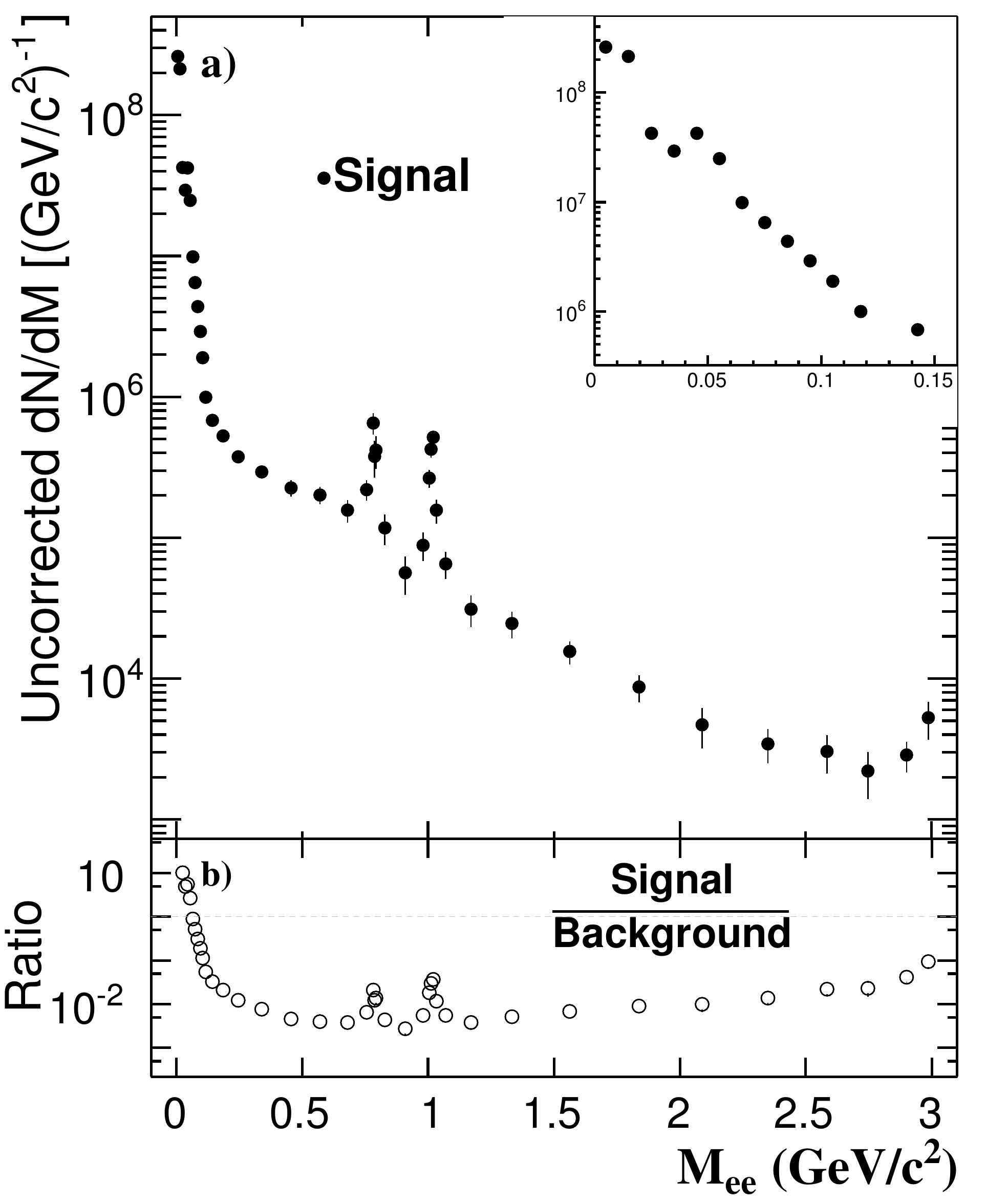}
\caption{Panel (a): The dielectron continuum
without efficiency correction in $\sqrt{s_{_{NN}}} = 200$ GeV
minimum-bias Au+Au collisions. The two-peak structure, as shown in the insert, for $M_{ee}\!<\!0.12$ GeV/$c^2$ is due to photon conversions in the beam pipe and supporting structure.
Errors are statistical only. Panel
(b): The signal over background ratio in $\sqrt{s_{_{NN}}} = 200$
GeV minimum-bias Au+Au collisions. The first two data points are not shown for clarity. Errors are statistical.}
\label{rawcontinuum}
\end{center}
\end{figure}
In the mixed-event technique, tracks from different events are used to form unlike-sign or like-sign pairs. The events are divided into 9000
categories according to the collision vertex (10 bins), event plane (defined in Sect.~\ref{flowmethod})
azimuthal angle (100 bins from 0 to $\pi/2$), and centrality (9 bins). The two events to be mixed must come
from the same event category to ensure  similar detector geometric acceptance,
azimuthal anisotropy, and track multiplicities. We find that when the number of event plane bins is larger than or equal to 30, the mixed-event spectrum describes the
combinatorial background.

In the same-event like-sign technique, electrons with the same
charge sign from the same events are paired. Due to the sector
structure of the TPC, and the different bending directions of positively
and negatively charged particle tracks in the transverse plane, like-sign and unlike-sign pairs have
different acceptances. The correction for this acceptance difference is applied to
the same-event like-sign pair distribution before background
subtraction. The acceptance difference between same-event unlike-sign and
same-event like-sign pairs is obtained using the mixed-event
technique. Fig.~\ref{accepplot} (a) shows the mixed-event
unlike-sign and mixed-event like-sign electron pair invariant mass
distributions in $\sqrt{s_{_{NN}}} = 200$ GeV minimum-bias Au+Au
collisions. The ratio of these two distributions, the acceptance difference factor, is shown in
Fig.~\ref{accepplot} (b), and its zoom-in version is shown in Fig.~\ref{accepplot} (c). 
The centrality and $p_T$ dependences are presented in Figs.~\ref{accepratplot} and~\ref{accepratptplot}, respectively. These figures show that the acceptance differences at low invariant mass are largest at low $p_T$ and in the most central collisions.

After correcting for the acceptance
difference, the same-event like-sign distribution is compared to the same-event unlike-sign pair distribution (which contains the signal) and the mixed-event unlike-sign pair distribution in
Fig.~\ref{eepair}~(a).  The
mixed-event unlike-sign distribution is normalized to
match the same-event like-sign distribution in the mass region
$0.9$$-$$3.0$~GeV/$c^{2}$. For $M_{ee}\!>\!0.9$ GeV/$c^{2}$, the ratio of the
same-event like-sign over the normalized mixed-event unlike-sign distributions is found constant with $\chi^{2}/NDF$ of 15/16, as shown in Fig.~\ref{eepair} (b). The constant is $0.9999\pm0.0004$.
The zoom-in version, centrality dependence, and $p_T$ dependence of this ratio are shown in Figs.~\ref{eepair} (c), ~\ref{centrat}, and~\ref{ptrat}, respectively. In addition, the centrality and $p_T$ dependences of the ratio of the same-event like-sign over the normalized mixed-event like-sign distributions are presented in Figs.~\ref{likesigncentrat} and~\ref{likesignptrat}, respectively.

In the low mass region, the correlated cross-pair background is present in the same-event like-sign
distribution, but not in the mixed-event unlike-sign background. In
the higher mass region, the mixed-event unlike-sign distribution
matches the same-event like-sign distribution. Therefore, for
$M_{ee}\!<\!0.9$ GeV/$c^{2}$ like-sign pairs from the same events
are used for background subtraction. For $M_{ee}\!>\!0.9$
GeV/$c^{2}$ we subtract the mixed-event unlike-sign background
to achieve better statistical precision.

Figure~\ref{2drawcontinuum} shows the $p_T$ as a function of $M_{ee}$ for dielectron continuum after background subtraction 
without efficiency correction in $\sqrt{s_{_{NN}}} = 200$ GeV
minimum-bias Au+Au collisions. Figure~\ref{rawcontinuum} (a) shows the dielectron-signal mass distribution in minimum-bias Au+Au collisions at
$\sqrt{s_{_{NN}}} = 200$ GeV. The analysis requires
$|y_{e^{+}e^{-}}|\!<1, |\eta_{e}|\!<1$ and $p_T(e)\!>0.2$ GeV/$c$. The distribution is not corrected for efficiency.
The signal-to-background (S/B) ratio in Au+Au collisions versus $M_{ee}$ is shown in
Fig.~\ref{rawcontinuum}~(b).

\subsection{Method to obtain azimuthal anisotropy}\label{flowmethod}
\renewcommand{\floatpagefraction}{0.75}
\begin{figure}[htbp]
\begin{center}
\includegraphics[keepaspectratio,width=0.49\textwidth]{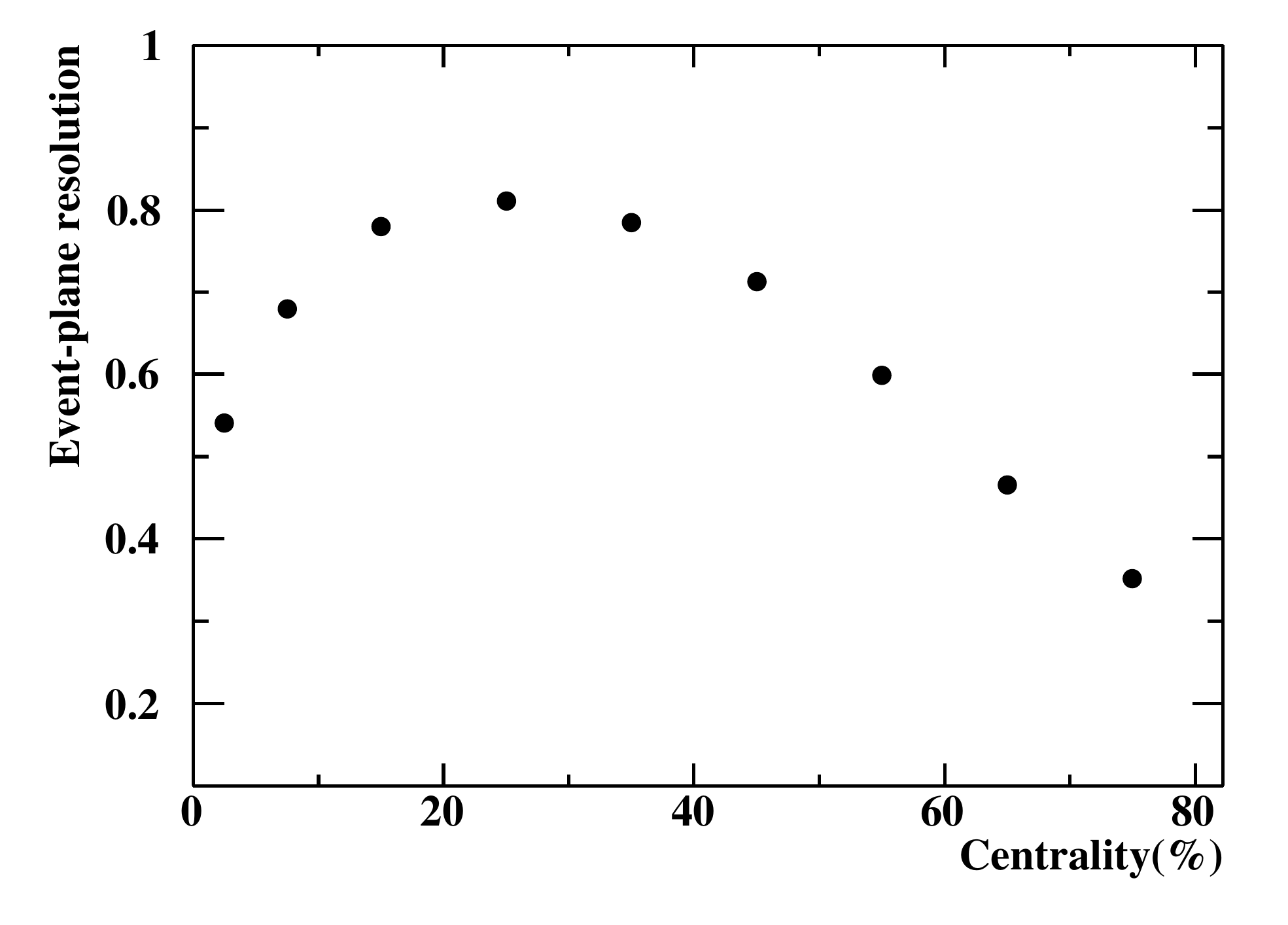}
\caption{The event-plane resolution from central to
peripheral (left to right) collisions in minimum-bias Au+Au collisions at
$\sqrt{s_{_{NN}}} = 200$ GeV.} \label{epreso}
\end{center}
\end{figure}

Hydrodynamic flow of  produced particles leads to azimuthal correlations 
among particles relative to the reaction plane~\cite{Art:98}.
However, the measured correlations also include effects not related to reaction plane orientation. These
are usually referred to as non-flow, and are due to, for example, 
resonance decays and parton fragmentation. In this analysis, we use the
'event-plane' method to determine the azimuthal anisotropy of produced dielectrons~\cite{Art:98}.
\renewcommand{\floatpagefraction}{0.75}
\begin{figure}[htbp]
\begin{center}
\includegraphics[keepaspectratio,width=0.49\textwidth]{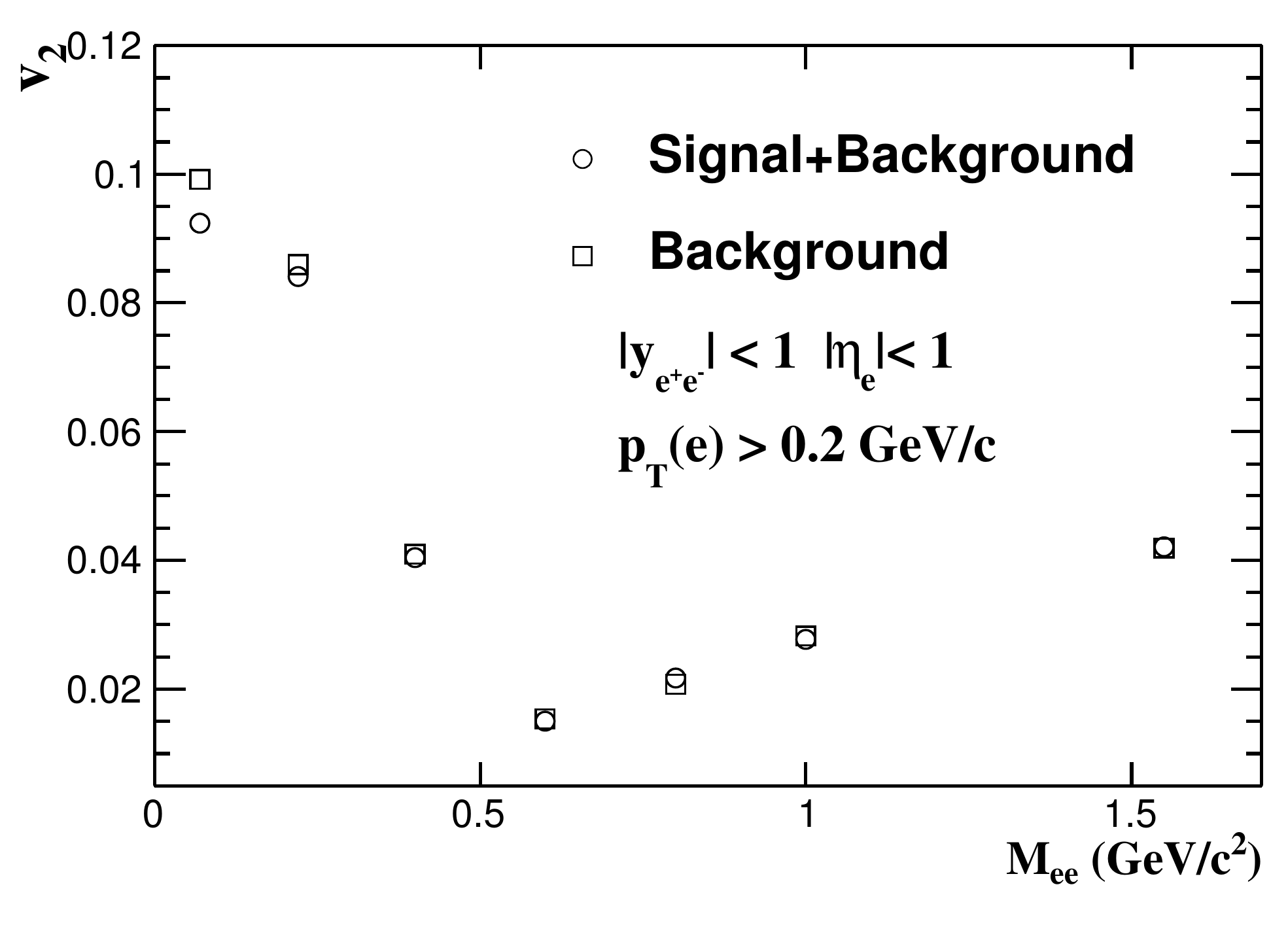}
\caption{The $v_2$ for the same-event
unlike-sign electron pairs (circles) and background (squares) as
a function of $M_{ee}$ within STAR acceptance in minimum-bias Au+Au collisions at
$\sqrt{s_{_{NN}}} = 200$ GeV. }
\label{v2method}
\end{center}
\end{figure}

The event plane is reconstructed using  tracks from the TPC.
The event flow vector $Q_2$ and the event-plane angle $\Psi_{2}$
are defined by~\cite{Art:98}:

\begin{equation}
Q_{2}\cos(2\Psi_{2})=Q_{2x}=\sum_{i}w_{i}\cos(2\phi_{i})
\end{equation}

\begin{equation}
Q_{2}\sin(2\Psi_{2})=Q_{2y}=\sum_{i}w_{i}\sin(2\phi_{i})
\end{equation}

\begin{equation}
\Psi_{2}= \bigg{(}\tan^{-1}\frac{Q_{2y}}{Q_{2x}}\bigg{)}/2,
\end{equation} where the summation is over all particles $i$ used for event-plane determination. Here, $\phi_{i}$ and $w_{i}$ are measured azimuthal angle and weight
for the particle $i$, respectively. The weight $w_{i}$ is equal to the particle $p_T$ up to 2 GeV/$c$, and is kept constant at higher $p_T$.  The electron candidates are excluded in the event-plane reconstruction to avoid the self-correlation effect. A PYTHIA study indicates that decay kaons from heavy flavor have no additional effect on event-plane determination.  An azimuthally non-homogeneous acceptance or efficiency of the detectors can introduce a bias in the event-plane reconstruction which would result in a non-uniform $\Psi_{2}$ angle distribution in the laboratory coordinate system. The recentering and shifting methods~\cite{flattenmethod,starv2:13} were used to flatten the $\Psi_{2}$ distribution.

The observed $v_2$ is the second harmonic of the azimuthal
distribution of particles with respect to the event plane:
\begin{equation}
v_{2}^\mathrm{obs}= \langle \cos[2(\phi-\Psi_{2})]\rangle~~~,
\end{equation}
where angle brackets denote an average over all particles with
azimuthal angle $\phi$ in a given phase space and $\phi-\Psi_{2}$ ranges from 0 to $\pi/2$. The electron reconstruction efficiency is independent of 
$\phi-\Psi_{2}$. The real
$v_2$ is corrected for event-plane resolution as

\begin{equation}
v_{2}=
\frac{v_{2}^\mathrm{obs}}{C\sqrt{{\langle\cos[2(\Psi_{2}^{a}-\Psi_{2}^{b})]\rangle}}}~~~,
\end{equation}
where $\Psi_{2}^{a}$ and
$\Psi_{2}^{b}$ are the second-order event planes determined from different sub-events, $C$ is a constant calculated from the known multiplicity dependence of the resolution~\cite{Art:98}, and the brackets denote an average over a large event sample. The denominator
represents the event-plane resolution, which is obtained from two
random sub-events~\cite{ksv2}. Figure~\ref{epreso} shows the
event-plane resolution for different centralities in 200 GeV Au+Au
collisions.

The $v_2$ for dielectron signals for each mass and $p_T$ bin is
obtained using the formula

\begin{equation}
v_{2}^{S}(M_{ee},p_{T}) =  \frac{v_{2}^\mathrm{total}(M_{ee},p_{T})}{r(M_{ee},p_{T})} 
-\frac{1-r(M_{ee},p_{T})}{r(M_{ee},p_{T})}v_{2}^{B}(M_{ee},p_{T}),
\end{equation}
in which $v_{2}^{S}$, $v_{2}^\mathrm{total}$, and $v_{2}^{B}$ represent $v_2$ for the dielectron signal, $v_2$ for the same-event
unlike-sign electron pairs, and $v_2$ for the
background electron pairs (determined through either the mixed-event
unlike-sign technique or the same-event like-sign method, as discussed
in the previous sections), respectively.  The parameter 
$r$ represents the ratio of the number of
dielectron signals ($N_{S}$) to the number of the same-event unlike-sign electron pairs ($N_{S+B}$).
The $v_{2}^\mathrm{total}$ is the yield-weighted average from the dielectron signal and background.
The mixed-event unlike-sign technique is applied for $M_{ee}\!>\!0.9$ GeV/$c^2$, for which the mixed-event unlike-sign distribution for each of the  ($\phi-\Psi_{2}$) bins (the bin width is $\frac{\pi}{10}$) is normalized to the corresponding same-event like-sign distribution in the same $\phi-\Psi_{2}$ bin. For the five ($\phi-\Psi_{2}$) bins, the normalization factors differ by 0.1\%. 
Figure~\ref{v2method} shows $v_{2}^\mathrm{total}$ and
$v_{2}^{B}$ as a function of $M_{ee}$ within the STAR acceptance in minimum-bias Au+Au
collisions at $\sqrt{s_{_{NN}}} = 200$ GeV. 
\subsection{Cocktail simulation}\label{simu}
\renewcommand{\floatpagefraction}{0.75}
\begin{figure*}[htbp]
\begin{center}
\includegraphics[keepaspectratio,width=0.85\textwidth]{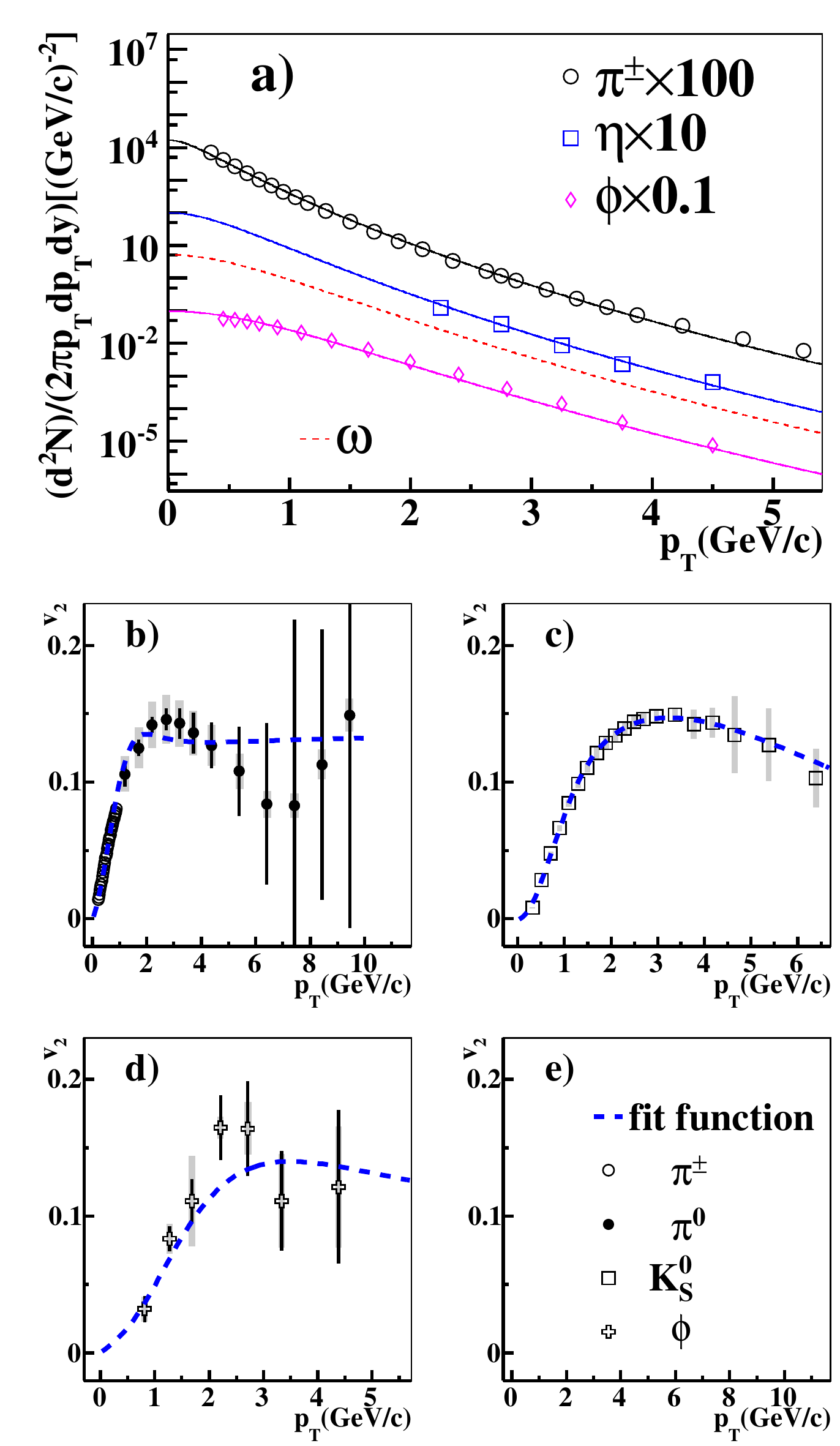}
\caption{(Color online) Panel (a): The invariant yields of
identified mesons, fit with Tsallis functions~\cite{Tsallis} in Au+Au
collisions at $\sqrt{s_{NN}} = 200$ GeV. See text for details.
Panels (b-d): The $v_2$ of
identified mesons, fit with a function
for $\sqrt{s_{NN}} = 200$ GeV minimum-bias Au+Au collisions. See text for details.} \label{spectrumv2input}
\end{center}
\end{figure*}
In the following we wish to obtain a representation of the dielectron $v_2$ distributions in $p_T$ and $M_{ee}$ by a cocktail simulation that accounts for the decays of all prominent hadronic sources. We shall obtain the dielectron $v_2$ from each decay component by combining the measured $p_T$ spectra of the ``mother mesons", with the previously measured $v_2$ distributions of these mesons. 

As mentioned earlier, the dielectron
pairs may come from decays of  light-flavor and heavy-flavor
hadrons. Contributions from the following hadronic sources and processes were included in the 
cocktail simulation to compare with the measured data: $\pi^{0}\rightarrow \gamma e^{+}e^{-}$, $\eta
\rightarrow \gamma e^{+}e^{-}$, $\omega \rightarrow
\pi^{0} e^{+}e^{-}$, $\omega \rightarrow e^{+}e^{-}$, $\phi \rightarrow \eta e^{+}e^{-}$, and $\phi
\rightarrow e^{+}e^{-}$ for $M_{ee}\!<$ 1.1 GeV/$c^2$. In the intermediate mass region, we simulate the dielectron $v_2$ from the $c\bar{c}$ correlated contribution.

\renewcommand{\floatpagefraction}{0.75}
\begin{figure}[htbp]
\begin{center}
\includegraphics[keepaspectratio,width=0.49\textwidth]{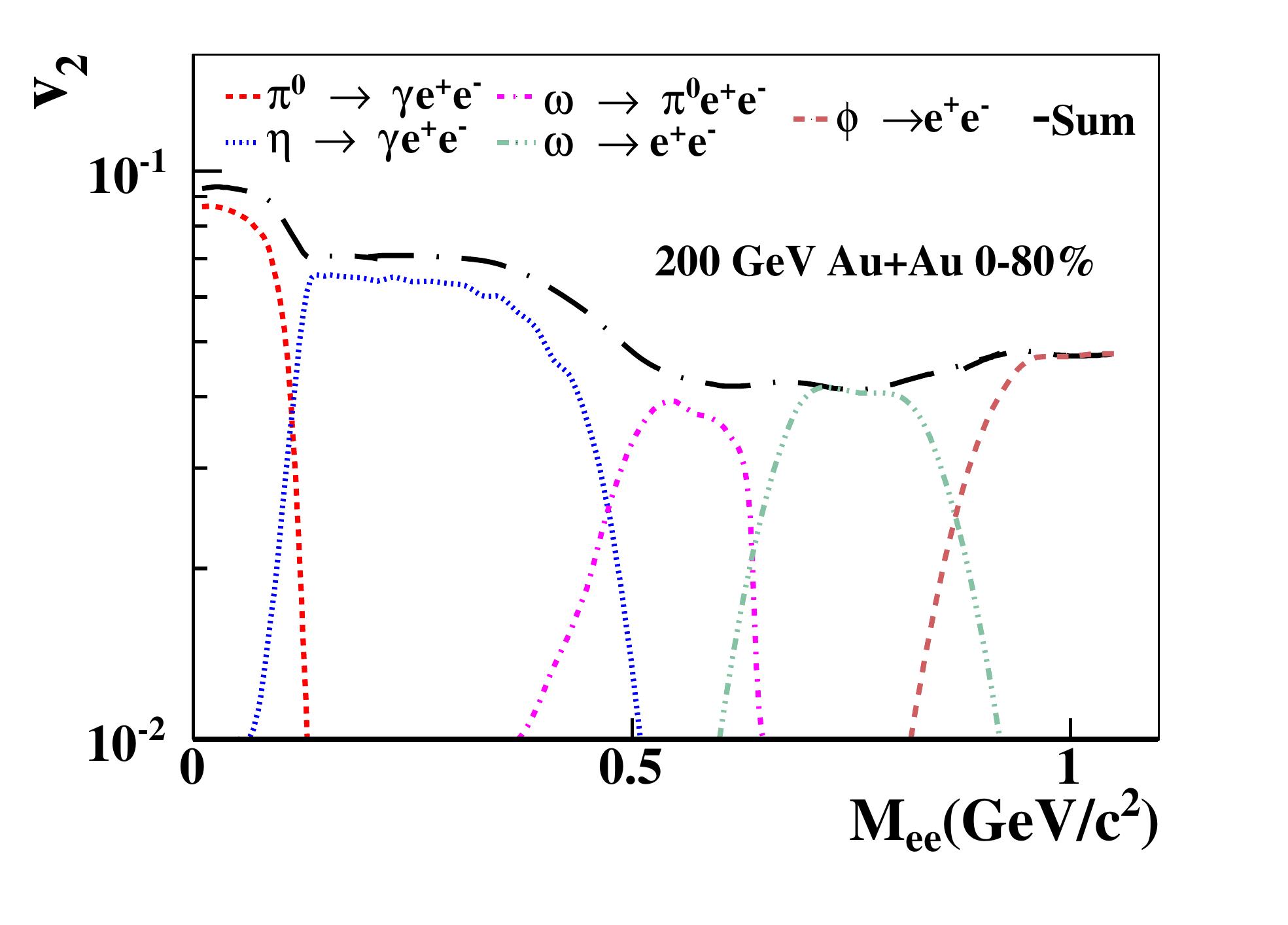}
\caption{(Color online) The simulated $v_2$ as a function of $M_{ee}$ from $\pi^{0}$, $\eta$, $\omega$ and $\phi$ decays within the STAR acceptance in minimum-bias Au+Au collisions at $\sqrt{s_{_{NN}}}
= 200$ GeV, including the contributions from specific decays. The contribution from $\phi \rightarrow \eta e^{+}e^{-}$ is smaller than 1\% and is not shown for clarity. The bin width is 20 MeV/$c^2$.} \label{cocktailv2}
\end{center}
\end{figure}
The  $\pi^{0}$ invariant yield is taken as the average of
$\pi^{+}$ and $\pi^{-}$~\cite{AuAuPID,pidlowpt}. The
$\phi$ yield is taken from STAR measurements~\cite{starphi}, while the $\eta$ yield is from a PHENIX measurement~\cite{phenixeta}.
We fit the meson invariant yields with Tsallis
functions~\cite{Tsallis}, as shown in Fig.~\ref{spectrumv2input} (a). The $\omega$ $p_T$-spectrum shape is derived from the Tsallis function. The $\omega$ total yield at mid-rapidity ($dN/dy|_{y=0}$) is obtained by matching the simulated cocktail to the efficiency-corrected dielectron mass spectrum in the $\omega$ peak region.  
Table~\ref{tab:III} lists the $dN/dy|_{y=0}$ of hadrons in 200 GeV minimum-bias Au+Au collisions.
In addition, we parameterize the $\pi$, $K_{S}^{0}$ and $\phi$ $v_2$ from previous measurements~\cite{piv2,starpiv2,ksv2,phiv2} with a data-driven functional form, 
$A\tanh(Bp_T)+C\mathrm{arctan}(Dp_T)+Ee^{-p_T}+Fe^{-p_T^{2}}$,where $A$, $B$, $C$, $D$, $E$, and $F$ are fit parameters.  The $\eta$ and $\omega$ $v_2$ are assumed to be the same as $K_{S}^{0}$ and $\phi$ $v_2$ respectively, since the masses of the $\eta$ and $K_{S}^{0}$ mesons, as well as those of the $\omega$ and $\phi$ mesons, are similar. The mass-dependent hydrodynamic behavior was observed for hadron $v_2$ at $p_{T}\!<2$ GeV/$c$ while in the range of $2\!<\!p_T\!<\!6$ GeV/$c$, the number of constituent quark scaling was observed in Au+Au collisions at $\sqrt{s_{_{NN}}}
= 200$ GeV~\cite{starwhitepaper,otherwhitepapers}.  Due to different methods and detector configurations, the non-flow effects vary from 3-5\% for charged and neutral $\pi$ measured by PHENIX to 15-20\% for charged $\pi$, $K_{S}^{0}$, and $\phi$ measured by STAR.
Figures.~\ref{spectrumv2input} (b-d) show the previously measured meson $v_2$ and the fit functions.
\renewcommand{\floatpagefraction}{0.75}
\begin{figure}[htbp]
\begin{center}
\includegraphics[keepaspectratio,width=0.49\textwidth]{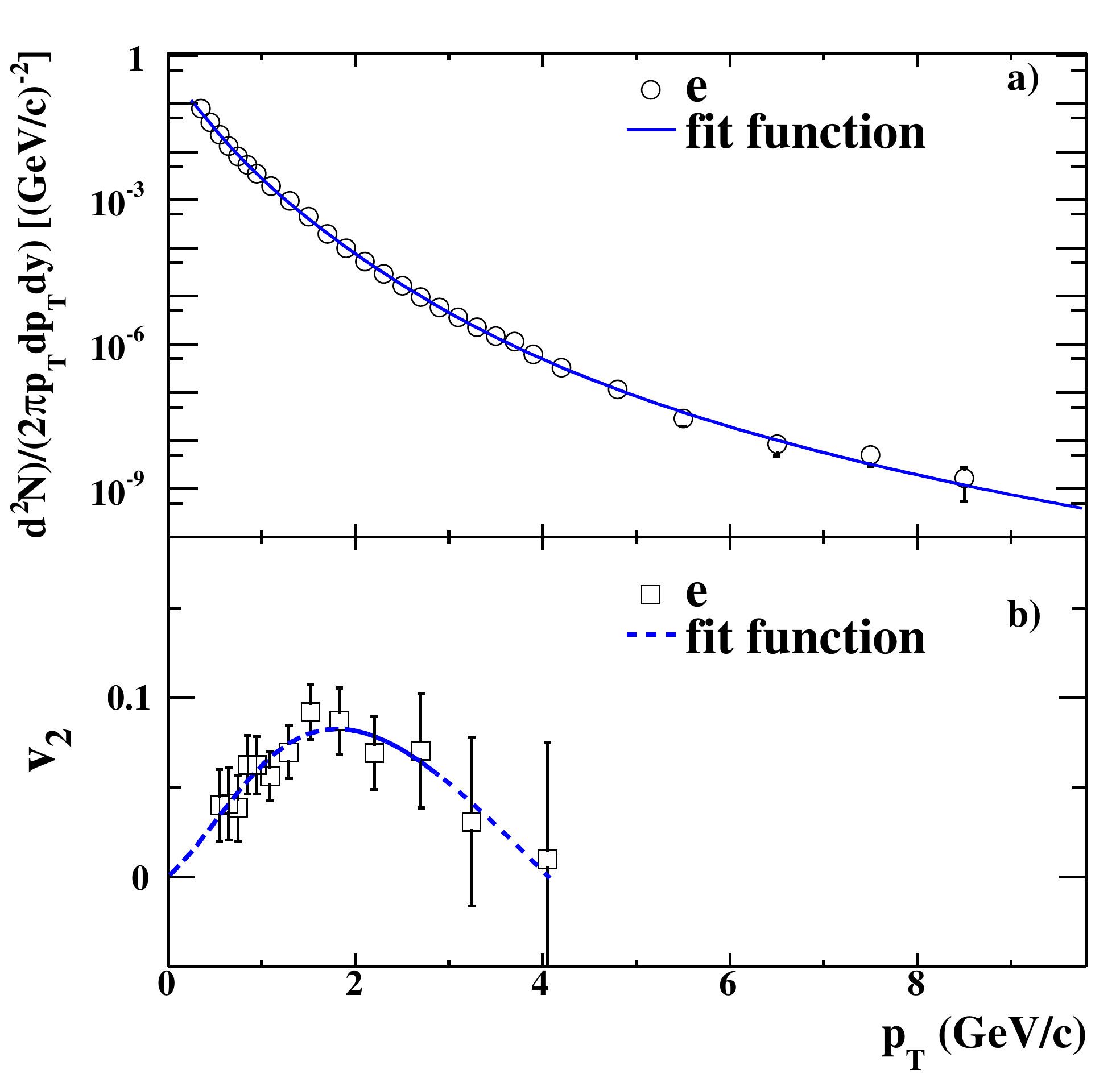}
\caption{(Color online) The invariant yield and $v_2$ of electrons from heavy flavor decays~\cite{PHENIXNPE}
fitted with functions in minimum-bias Au+Au collisions at $\sqrt{s_{_{NN}}}
= 200$ GeV. The spectrum is fit with a function $A[e^{B\sqrt{{p_{T}}^2+C}+D({p_{T}}^2+C)}+\sqrt{{p_{T}}^2+C}/E]^{F}$, where $A$, $B$, $C$, $D$, $E$, and $F$ are fit parameters. The $v_2$ is fit with the same function as used to parameterize the meson $v_2$ shown in Fig.\ref{spectrumv2input}. } \label{electronv2input}
\end{center}
\end{figure}

\begin{table}\caption{The total yields at mid-rapidity ($dN/dy$) from the Tsallis fit and
decay branching ratios of hadrons in minimum-bias Au+Au collisions at $\sqrt{s_{_{NN}}}
= 200$ GeV.\label{tab:III}}
{\centering
\begin{tabular}{c|c|c|c|c} \hline\hline
 meson & $\frac{dN}{dy}$ & relative uncertainty & decay channel & BR \\
\hline
$\pi^{0}$   & $98.5$ & 8\% & $\gamma e^{+}e^{-}$ & $1.174\times10^{-2}$ \\
$\eta$   & $7.86$ & 30\% & $\gamma e^{+}e^{-}$ & \hspace{0.14in}$7.0\times10^{-3}$ \\
$\omega$   & $9.87$ & 33\% & $e^{+}e^{-}$ & \hspace{0.065in}$7.28\times10^{-5}$ \\
$\omega$   & & & $\pi^{0}e^{+}e^{-}$ & \hspace{0.14in}$7.7\times10^{-4}$ \\
$\phi$   & $2.43$ & 10\% & $e^{+}e^{-}$ & $2.954\times10^{-4}$ \\
$\phi$   & & & $\eta e^{+}e^{-}$ & \hspace{0.065in}$1.15\times10^{-4}$ \\
\hline \hline
\end{tabular}
}
\end{table}

\renewcommand{\floatpagefraction}{0.75}
\begin{figure}[htbp]
\begin{center}
\includegraphics[keepaspectratio,width=0.49\textwidth]{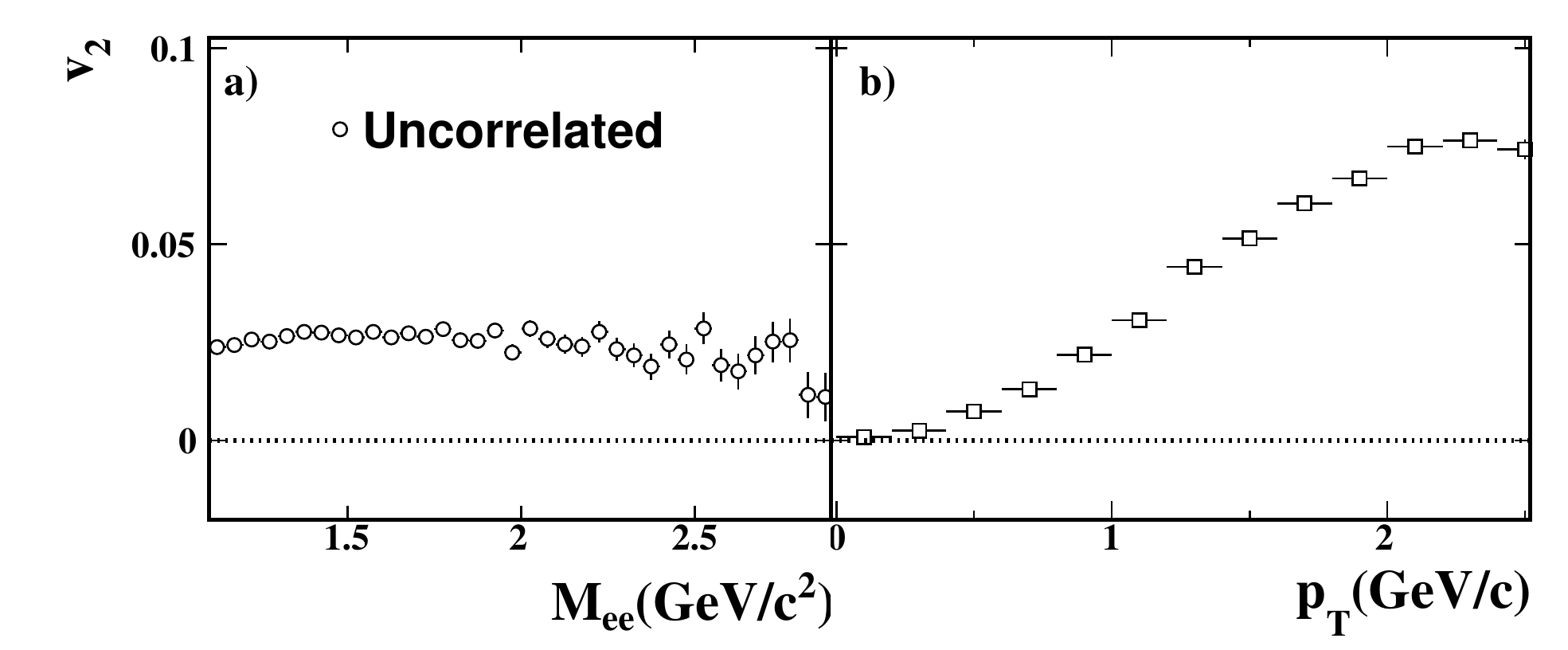}
\caption{The dielectron $v_2$ from the $c\bar{c}$ contribution as a function of $M_{ee}$ and $p_T$ with a completely uncorrelated $c$ and $\bar{c}$.} \label{charmv2}
\end{center}
\end{figure}
With the Tsallis functions for the spectra and the parameterizations for $v_2$ as input, we simulate decays of $\pi^{0}$, $\eta$, $\omega$ and $\phi$ with appropriate branching ratios (BRs), and obtain the dielectron $v_2$, as shown in Fig.~\ref{cocktailv2}. The final $v_2$ is the yield-weighted average from different contributions. The same acceptance conditions after momentum resolution smearing  are utilized as those used in the analysis of real events. The Kroll-Wada expression is used for the Dalitz decay: $\pi^{0}\rightarrow \gamma e^{+}e^{-}$,
$\eta \rightarrow \gamma e^{+}e^{-}$, $\omega \rightarrow
\pi^{0} e^{+}e^{-}$ and $\phi \rightarrow \eta e^{+}e^{-}$~\cite{starppdilepton,krollwada,ruan:11}.

In different mass regions different particle species dominate the
production, as listed in Table~\ref{tabII}~\cite{staromega,starppdilepton}.
Studying $v_2$ in different mass regions
should therefore help discern the azimuthal anisotropy
of different species.
Figure~\ref{cocktailv2} shows that among $\pi^{0}$, $\eta$, $\omega$ and $\phi$ decays, $\pi^{0}\rightarrow \gamma e^{+}e^{-}$, $\eta \rightarrow \gamma e^{+}e^{-}$, $\omega \rightarrow
\pi^{0} e^{+}e^{-}$, $\omega \rightarrow e^{+}e^{-}$, and $\phi \rightarrow e^{+}e^{-}$ dominate the $v_2$ contribution in the mass regions $[0, 0.14]$, $[0.14, 0.30]$, $[0.5, 0.7]$,$[0.76, 0.80]$, and $[0.98, 1.06]$ GeV/$c^2$ respectively.
\begin{table}\caption{The sources of
dielectrons in different mass regions. \label{tabII}}
{\centering
\begin{tabular}{c|c} \hline\hline
 mass region & dominant source(s)\\
 (GeV/$c^2$)& of dielectrons\\ \hline
\hspace{0.16in}0--0.14&  $\pi^{0}$ and photon conversions\\
0.14--0.30 & $\eta$\\
0.50--0.70 & $\textrm{charm}+\rho^0$ (in-medium) \\
 0.76--0.80 & $\omega$ \\
0.98--1.06 & $\phi$ \\
1.1--2.9 & $\textrm{charm}$ + thermal radiation \\
\hline \hline
\end{tabular}
}
\end{table}

For $1.1\!<M_{ee}\!<2.9$  GeV/$c^2$, we simulate the dielectron $v_2$  from $c\bar{c}$ correlated contributions.
To get a handle on the unknown $c\bar{c} \rightarrow e^{+}e^{-}X$ correlation in Au+Au collisions, we take two extreme approaches to simulate this $v_2$ contribution: 1) we assume the $c$ and $\bar{c}$ are completely uncorrelated; 2) we assume the $c$ and $\bar{c}$ correlation is the same as shown in PYTHIA 6.416, in
which the $k_T$ factor is set by PARP(91)=1 GeV/c, and the parton
shower is set by PARP(67)=1~\cite{pythia}. With these parameter
values, PYTHIA can describe the shape of the $D^{0}$~\cite{Yifei:2012}
spectrum and the non-photonic electron spectrum measured by STAR ~\cite{starelectron,wei:2011} for $p+p$ collisions.

In Fig.~\ref{electronv2input}, the measured spectrum and $v_2$ of electrons from heavy-flavor decays~\cite{PHENIXNPE} are shown as well as results of a parameterization which is used to obtain the dielectron $v_2$ from the $c\bar{c}$ contribution. We find the dielectron $v_2$ from $c\bar{c}$ contribution does not show a significant difference for the two cases explained above. The $v_2$ value is 0.022 for the PYTHIA-correlation case and 0.027 for the uncorrelated case. Therefore, in the subsequent sections, we use the uncorrelated result to compare with our measurements. Figure~\ref{charmv2} shows the dielectron $v_2$ from the $c\bar{c}$ contribution as a function of $M_{ee}$ and $p_T$ with  a completely uncorrelated $c$ and $\bar{c}$.  
\renewcommand{\floatpagefraction}{0.75}
\begin{figure*}[htbp]
\begin{center}
\includegraphics[keepaspectratio,width=0.99\textwidth]{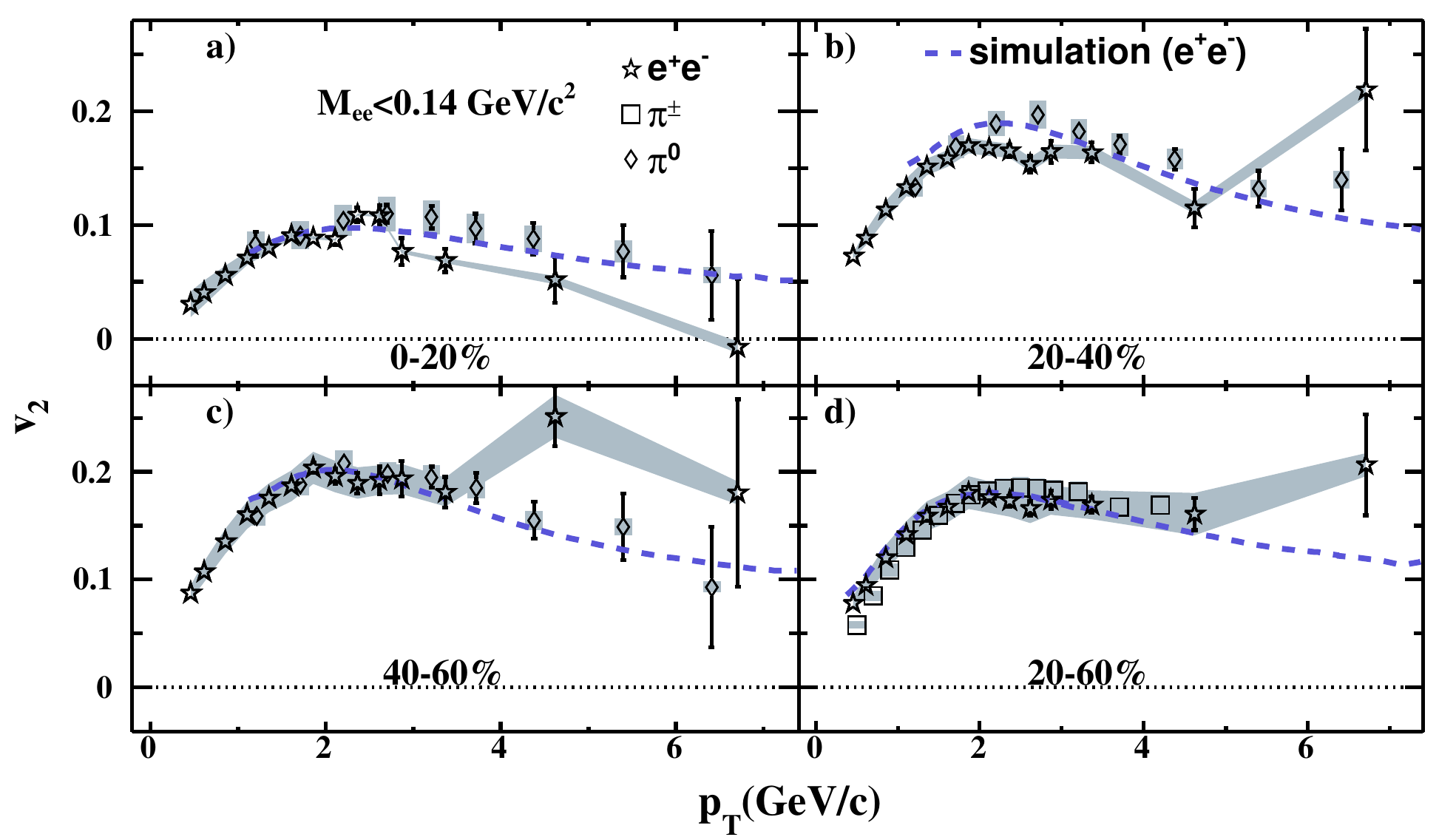}
\caption{(Color online) The dielectron $v_2$ in the $\pi^{0}$
Dalitz decay region (star symbol) as a function of $p_T$ in
different centralities from Au+Au collisions at $\sqrt{s_{_{NN}}}
= 200$ GeV. Also shown are the charged (square)~\cite{chargepiv2}, neutral (diamond) ~\cite{piv2} pion $v_2$, and
the expected dielectron $v_2$ (dashed curve) from $\pi^{0}$
Dalitz decay. The bars and bands represent statistical and
systematic uncertainties, respectively.} \label{piv2}
\end{center}
\end{figure*}

\section{Systematic uncertainties}\label{sys}
\renewcommand{\floatpagefraction}{0.75}
\begin{figure*}[htbp]
\begin{center}
\includegraphics[keepaspectratio,width=0.99\textwidth]{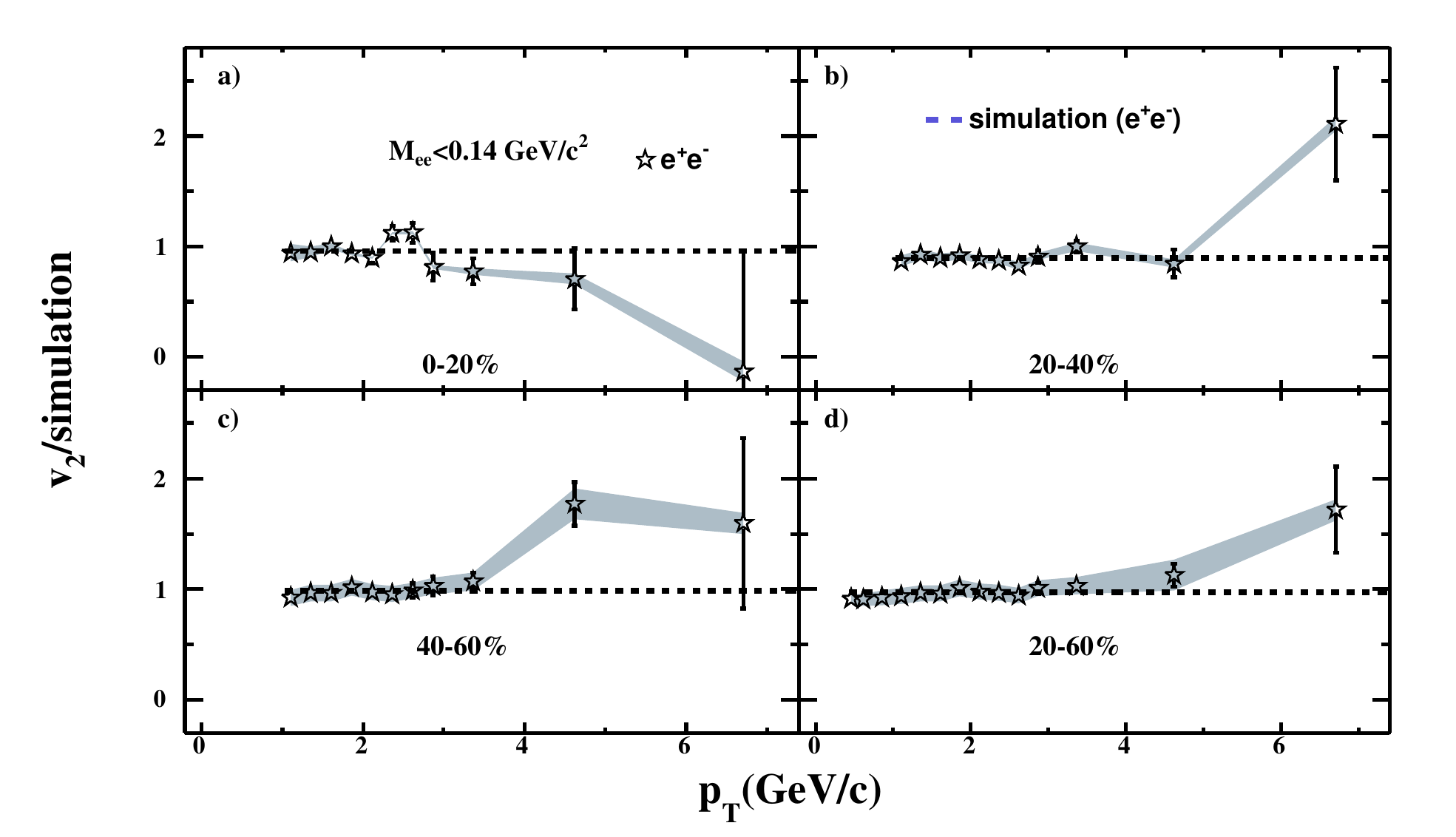}
\caption{(Color online) The ratio of the measured dielectron $v_2$ in the $\pi^{0}$
Dalitz decay region over the expected dielectron $v_2$ from $\pi^{0}$
Dalitz decay as a function of $p_T$ in different centralities from Au+Au collisions at $\sqrt{s_{_{NN}}}
= 200$ GeV. The dashed line is a constant fit to the ratio. The bars and bands represent statistical and systematic uncertainties, respectively.} \label{piv2rat}
\end{center}
\end{figure*}

\begin{table*}\caption{Sources and their contributions to the absolute systematic uncertainties for dielectron $v_2$ measurements in different mass regions. The uncertainties for each source are $p_T$ dependent and listed as a range for each mass region. The total absolute systematic uncertainties are the quadratic sums of the different contributions. NR represents normalization range.
 \label{tab:IV}}
{\centering 
\begin{tabular}{c|c|c|c|c|c|c} \hline\hline
 source/ & $0-0.14$ & $0.14-0.30$ & $0.5-0.7$ &$0.76-0.80$ & $0.98-1.06$ & $1.1-2.9$ GeV/$c^2$\\
 contribution &  &  &  & & & \\
\hline
DCA cut & $(0.2-1.3)\times10^{-3}$& $(0.6-2.8)\times10^{-2}$ &$\hspace{0.1in}(2.5-9.7)\times10^{-2}$ & $(0.4-3.7)\times10^{-2}$&$(1.3-2.7)\times10^{-2}$ & $(0.9-12.5)\times10^{-2}$\\
NR & -- & -- &-- &-- & $(1.0-3.0)\times10^{-2}$& \hspace{0.05in}$(3.2-6.8)\times10^{-2}$\\
bg method & -- & -- &-- &-- &-- &  $-(8.0-34.1)\times10^{-2}$\\
$n\sigma_{e}$ cut &$\hspace{0.2in}<1\times10^{-4}$ & $(0.1-0.4)\times10^{-2}$& $\hspace{0.1in}(0.2-0.4)\times10^{-2}$ &$(0.1-0.6)\times10^{-2}$ & $(0.3-1.0)\times10^{-2}$& $\hspace{0.05in}(0.3-2.8)\times10^{-2}$\\
$\eta-$gap & $(0.1-7.3)\times10^{-3}$ &-- &-- &-- &-- &-- \\
\hline
total & $(0.2-7.4)\times10^{-3}$ & $(0.6-2.8)\times10^{-2}$ &$(2.6-9.7)\times10^{-2}$ & $(0.5-3.7)\times10^{-2}$  &$(2.6-3.3)\times10^{-2}$ & $+(4.9-13.0)\times10^{-2}$\\
 &  &  & &   & & $-(9.4-36.5)\times10^{-2}$\\
\hline \hline
\end{tabular}
}
\end{table*}
The systematic uncertainties for the dielectron $v_2$ are dominated by
background subtraction. The combinatorial background effect is evaluated
by changing the DCA cut of the electron candidates.
We vary the DCA cut from less than 1 cm to less than 0.8 cm so that the number of dielectron pairs changes by 20\%.  

The uncertainties in the correction of the acceptance difference between same-event unlike-sign and same-event like-sign pairs are studied and found to have a negligible contribution. 

For $0.9\!<M_{ee}\!<2.9$ GeV/$c^2$, there are additional systematic uncertainties from the
mixed-event normalization and background subtraction methods. The
uncertainty on the mixed-event normalization is obtained by taking the full difference between the results from varying the normalization range from $0.9\!<M_{ee}\!<3.0$ to $0.7\!<M_{ee}\!<3.0$ GeV/$c^2$. In addition, there can be correlated sources in the same-event like-sign pairs for which the mixed-event background cannot completely account. This would lead to a larger $v_2$ for the dielectron signal when using mixed-event background subtraction. Therefore, the full difference between mixed-event unlike-sign and same-event like-sign background subtraction contributes to the lower bound of the systematic uncertainties.  In the mass region 0.98-1.06 GeV/$c^2$, the full difference between mixed-event unlike-sign and same-event like-sign background subtraction is negligible and not shown in Table~\ref{tab:IV}.

We also evaluate the hadron contamination effect by changing the
$n\sigma_{e}$ cut. The hadron contamination is varied from 5\% to 4\% and to 6\%. The $v_2$ difference between the default value and the new value is quoted as part of the systematic uncertainties, as shown in Table~\ref{tab:IV}. 
\renewcommand{\floatpagefraction}{0.75}
\begin{figure*}[htbp]
\begin{center}
\includegraphics[keepaspectratio,width=0.99\textwidth]{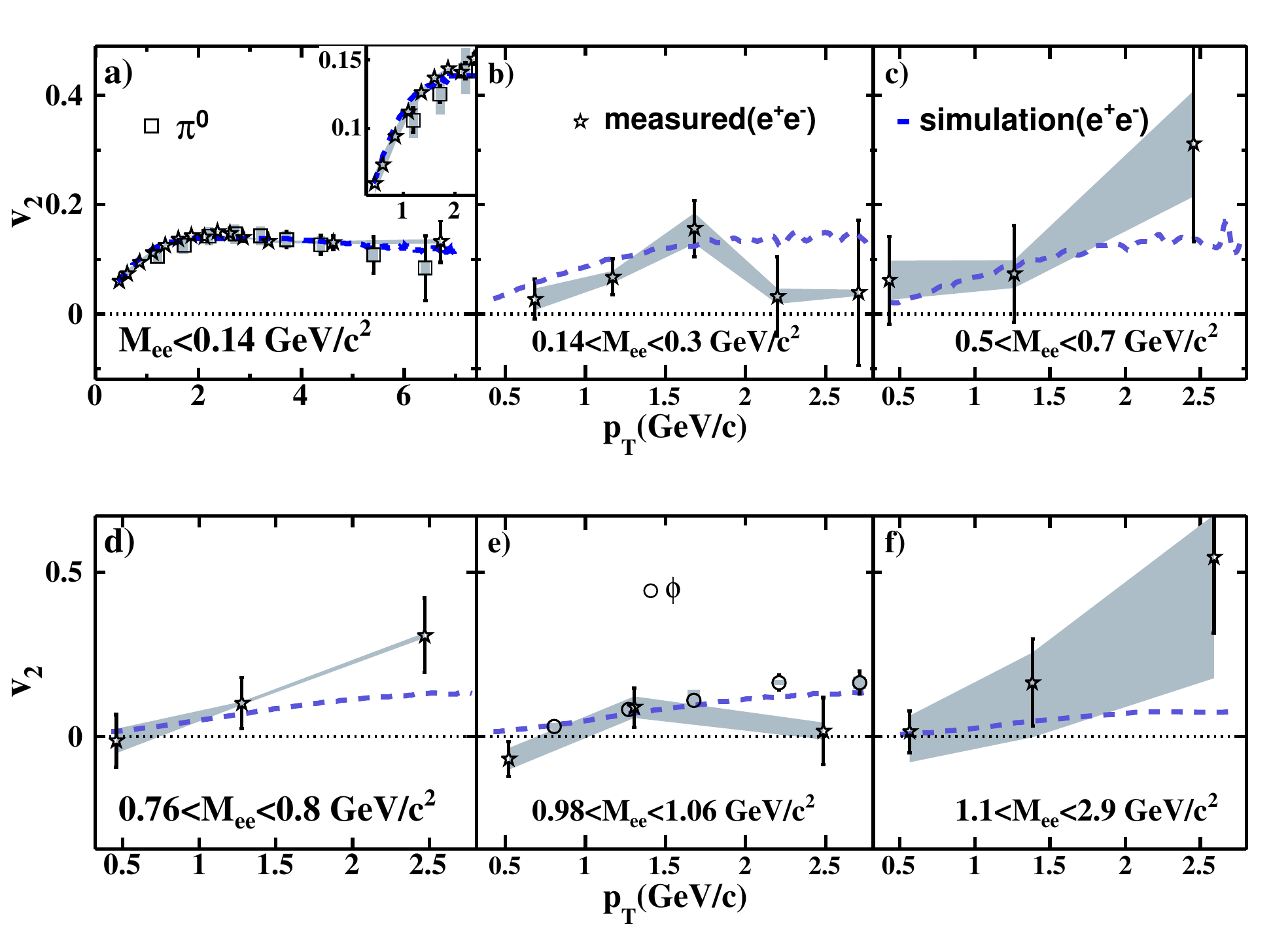}
\caption{(Color online) Panels (a-f): The dielectron $v_2$ as a function of $p_T$ in
minimum-bias Au+Au collisions at $\sqrt{s_{_{NN}}} = 200$ GeV for six different mass
regions: $\pi^{0}$, $\eta$, $\textrm{charm}+\rho^0$, $\omega$, $\phi$, and $\textrm{charm}$+thermal
radiation. Also shown are the neutral pion~\cite{piv2}
$v_2$ and the $\phi$ meson $v_2$~\cite{phiv2} measured through the decay channel
$\phi\rightarrow K^{+}K^{-}$. The expected dielectron $v_2$ (dashed curves) from $\pi^{0}$, $\eta$, $\omega$ and $\phi$ decays in the relevant mass regions are shown in panels (a-e) while that from $c\bar{c}$ contributions is shown in panel (f). The bars and bands represent statistical
and systematic uncertainties, respectively. The full difference between mixed-event unlike-sign and same-event like-sign background subtraction contributes to the lower bound of the systematic uncertainties, which leads to asymmetric systematic uncertainties in panel (f).} \label{diffv2}
\end{center}
\end{figure*}

In addition, we use the $\eta-$subevent method~\cite{ksv2} to study
the systematic uncertainties for the dielectron $v_2$  in
the $\pi^{0}$ Dalitz decay mass region. An $\eta$ gap of
$|\eta|\!<0.3$ between positive and negative pseudorapidity
subevents is introduced to reduce non-flow
effects~\cite{ksv2}.  The $v_2$ difference between the $\eta-$subevent method and the default method contributes $(0.1-7.3)\times10^{-3}$ absolute systematic uncertainties for $M_{ee}\!<0.14$ GeV/$c^2$.
We do not study this effect for the dielectron $v_2$ in the other mass regions due to limited statistics. However, the systematic
uncertainty from this is expected to be much smaller than
the statistical precision of the dielectron $v_2$. 

The systematic uncertainties of dielectron $v_2$ for the 2010 and 2011 data sets are studied separately and found to be comparable. For the combined results, the systematic uncertainties are taken as the average from the two data sets. Table~\ref{tab:IV} lists sources and their contributions to the absolute systematic uncertainties for the dielectron $v_2$ values in different mass regions. For each mass region, the systematic uncertainties are $p_T$ dependent for each source. The total absolute systematic uncertainties are the quadratic sums of the different contributions.

\section{Results}\label{results}

\renewcommand{\floatpagefraction}{0.75}
\begin{figure}[htbp]
\begin{center}\includegraphics[keepaspectratio,width=0.49\textwidth]{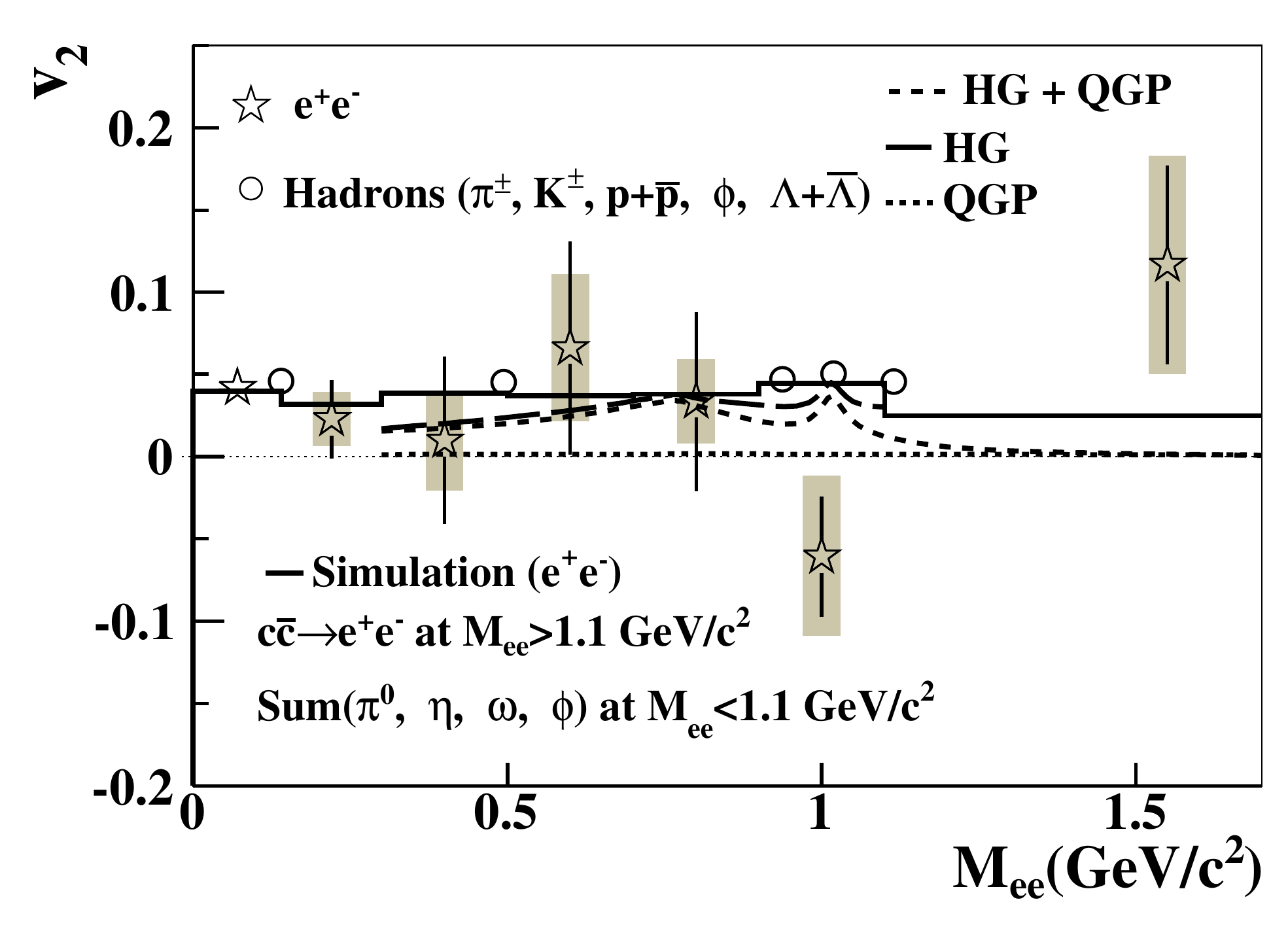}
\caption{(Color online) The $p_T$-integrated dielectron $v_2$ as a function of
$M_{ee}$ in minimum-bias Au+Au collisions at $\sqrt{s_{_{NN}}} = 200$ GeV. Also shown are the corresponding dielectron $v_2$ simulated from $\pi^{0}$, $\eta$, $\omega$ and $\phi$ decays and a $c\bar{c}$ contribution. The theoretical calculations from hadronic matter and QGP thermal radiation and the sum of these two sources~\cite{Gojko:12} are shown for comparisons. The $v_2$ for hadrons $\pi$, $K$, $p$, $\phi$, and $\Lambda$ are also shown for comparison. The bars and boxes represent statistical and systematic uncertainties, respectively. The systematic uncertainty for the first data point is smaller than the size of the marker.} \label{inclv2}
\end{center}
\end{figure}

The measured dielectron $v_2$ as a function of $p_T$ for $M_{ee}\!<0.14$ GeV/$c^2$ in different centralities from Au+Au collisions at $\sqrt{s_{_{NN}}}
= 200$ GeV are shown in Fig.~\ref{piv2}.  For comparison, the charged and neutral pion $v_2$ results~\cite{chargepiv2,piv2} are also shown in Fig.~\ref{piv2}. We parameterize the pion $v_2$ from low to high $p_T$, perform the
Dalitz decay simulation, and obtain the expected dielectron $v_2$
from $\pi^0$ Dalitz decay shown by the dashed curve. The ratio of the measured dielectron $v_2$ to the expected is presented in Fig.~\ref{piv2rat}. The simulated dielectron $v_2$  from $\pi^{0}$
Dalitz decay is consistent with our measurements in all
centralities within 5-10\%. We note that different non-flow effects in the dielectron $v_2$ analysis and the PHENIX $\pi$ $v_2$ analysis might contribute to differences between data and simulation.

Figure~\ref{diffv2} shows the dielectron $v_2$ as a function of $p_T$ in
minimum-bias Au+Au collisions at $\sqrt{s_{_{NN}}} = 200$ GeV in six different mass
regions: $\pi^{0}$, $\eta$, $\textrm{charm}+\rho^0$, $\omega$, $\phi$, and $\textrm{charm}$+thermal
radiation, as defined in Table~\ref{tabII}. We find that the expected dielectron $v_2$ (dashed curve) from $\pi^{0}$, $\eta$, $\omega$ and $\phi$ decays is consistent with the measured dielectron $v_2$ for $M_{ee}\!<1.1$ GeV/$c^2$. The dielectron $v_2$ in the $\phi$ mass region is consistent with the $\phi$ meson $v_2$ measured
through the decay channel $\phi\rightarrow K^{+}K^{-}$~\cite{phiv2}. In
addition, in the $\textrm{charm}$+thermal radiation mass region, dielectron $v_2$ can be described by a $c\bar{c}$ contribution within experimental uncertainties. 

With the measured $p_T$-differential $v_2$ presented above and cocktail spectrum shapes detailed in Sect.~\ref{simu}, we obtain the dielectron integral $v_2$ for $|y_{e^{+}e^{-}}|\!<1$, which is the yield weighted average for $p_T(e^{+}e^{-})\!>0$.  For the low $p_T$ region where the analysis is not applicable, we use the simulated differential $v_2$ for the extrapolation. The $p_T$ spectra of dielectrons might be different from those of cocktail components. For the mass region $0.2\! <\!M_{ee}\!< \!1.0$ GeV/$c^2$, we also use dielectron $p_T$ spectra measured by PHENIX~\cite{lowmass} and obtain the integral $v_2$ in these mass regions. The difference between this and the default case contributes additional systematic uncertainties
for the integral $v_2$ measurements, which are smaller than
those from other sources detailed in Sect.~\ref{sys}.
Figure~\ref{inclv2} shows the dielectron integral $v_2$ from data and simulation for $|y_{e^{+}e^{-}}|\!<1$ for minimum-bias Au+Au collisions at
$\sqrt{s_{_{NN}}} = 200$ GeV.  Also shown are the corresponding dielectron $v_2$ simulated from $\pi^{0}$, $\eta$, $\omega$, and $\phi$ decays and the $c\bar{c}$ contribution.

For $M_{ee}\!<1.1$  GeV/$c^2$, the $v_2$ from simulated $\pi^{0}$, $\eta$, $\omega$ and $\phi$ decays is consistent with the measured dielectron $v_2$ within experimental uncertainties. For the measured range  $1.1\!<M_{ee}\!<2.9$ GeV/$c^2$, the estimated $v_2$ magnitude from the simulated $c\bar{c}$ contribution is consistent with the measurement.

We also observe the measured dielectron integral $v_2$ as a function of $M_{ee}$ to be comparable to the hadron $v_2$ at a given hadron mass.  The hadron $v_2$ integral is obtained from the measured  $p_T$ differential $v_2$~\cite{phiv2,hadronv2,lambdav2} and spectrum shapes~\cite{Tsallis}. Also shown in Fig.~\ref{inclv2} is a comparison to theoretical calculations for the $v_2$ of thermally radiated dileptons from a hadron gas (HG) and the QGP separately, and for the sum of the two with a calculation of the relative contributions from HG and QGP~\cite{Gojko:12}. In this calculation, the dilepton $v_2$ are studied with 3+1D viscous hydrodynamics.  The QGP contribution comes from leading order quark-antiquark annihilation while for the HG emission rate, the Vector Dominance Model is used. According to this calculation, the dilepton radiation is QGP dominated for $M_{ee}\!>\!1.3$ GeV/$c^2$.  However, the charm $v_2$ must first be subtracted in order to compare directly with the theoretical calculation. In the future, with more data and more precise measurements of the charm contribution to the dielectron spectrum and $v_2$, hadron cocktail contributions may be subtracted from the measurements and the $v_2$ of excess dielectrons may be obtained.  The excess dielectron spectrum and  $v_2$ measurements as a function of $p_T$ in the mass region $1.3\!<M_{ee}\!<\!2.9$ GeV/$c^2$ will enable a direct comparison to theoretical results for QGP thermal radiation~\cite{Gojko:12}.

\section{Summary}\label{summary}
In summary, we report the first dielectron azimuthal anisotropy
measurement from Au+Au collisions at $\sqrt{s_{_{NN}}} = 200$ GeV.
The dielectron $v_2$  for $M_{ee}\!<1.1$ GeV/$c^2$ as a function of $p_T$ is found to be consistent with the
$v_2$ for $\pi^{0}$, $\eta$, $\omega$, and $\phi$ decays. For $1.1\!<M_{ee}\!<2.9$ GeV/$c^2$, the measured dielectron $v_2$ is described by the $c\bar{c}$ contribution within statistical and systematic uncertainties. With more data taken in the future, STAR will be in a good position to distinguish a QGP-dominated scenario from a HG-dominated one.

We thank C. Gale, R. Rapp, G. Vujanovic, and C. Young for valuable discussions and for
providing the theoretical calculations. We thank the RHIC Operations Group and RCF at BNL, the NERSC Center at LBNL, the KISTI Center in Korea and the Open Science Grid consortium for providing resources and support. This work was supported in part by the Offices of NP and HEP within the U.S. DOE Office of Science, the U.S. NSF, CNRS/IN2P3, FAPESP CNPq of Brazil, Ministry of Ed. and Sci. of the Russian Federation, NNSFC, CAS, MoST and MoE of China, the Korean Research Foundation, GA and MSMT of the Czech Republic, FIAS of Germany, DAE, DST, and CSIR of India, National Science Centre of Poland, National Research Foundation (NRF-2012004024), Ministry of Sci., Ed. and Sports of the Rep. of Croatia, and RosAtom of Russia.


\begin{thebibliography}{9}



\expandafter\ifx\csname
natexlab\endcsname\relax\def\natexlab#1{#1}\fi
\expandafter\ifx\csname bibnamefont\endcsname\relax
  \def\bibnamefont#1{#1}\fi
\expandafter\ifx\csname bibfnamefont\endcsname\relax
  \def\bibfnamefont#1{#1}\fi
\expandafter\ifx\csname citenamefont\endcsname\relax
  \def\citenamefont#1{#1}\fi
\expandafter\ifx\csname url\endcsname\relax
  \def\url#1{\texttt{#1}}\fi
\expandafter\ifx\csname
urlprefix\endcsname\relax\def\urlprefix{URL }\fi
\providecommand{\bibinfo}[2]{#2}
\providecommand{\eprint}[2][]{\url{#2}}

\bibitem{starwhitepaper} J. Adams {\it et al.},
\Journal{\NPA}{757}{102}{2005}.
\bibitem{otherwhitepapers} I. Arsene {\it et al.}, \Journal{\NPA}{757}{1}{2005}; K. Adcox
{\it et al.}, \Journal{\NPA}{757}{184}{2005}; B.B. Back {\it et
al.}, \Journal{\NPA}{757}{28}{2005}.
\bibitem{dilepton} R. Rapp and J. Wambach, Adv. Nucl. Phys. \textbf{25}, 1 (2000).
\bibitem{dileptonII} G. David, R. Rapp and Z. Xu, \Journal{\PR}{462}{176}{2008}.
\bibitem{ceres}G. Agakichiev {\it et al.}, \Journal{\EPJ}{41}{475}{2005}.
\bibitem{na60} R. Arnaldi {\it et al.}, \Journal{\PRL}{96}{162302}{2006}.
\bibitem{dropmass} G.E. Brown and M. Rho, Phys. Rep. \textbf{269}, 333 (1996).
\bibitem{massbroaden} R. Rapp and J. Wambach, Eur. Phys. J. A \textbf{6}, 415
(1999). 
\bibitem{massbroadenII}
K. Dusling, D. Teaney and I. Zahed, \Journal{\PRC}{75}{024908}{2007};   
H. van Hees and R. Rapp, \Journal{\NPA}{806}{339}{2008};
T. Renk and J. Ruppert, \Journal{\PRC}{77}{024907}{2008}.
\bibitem{lowmass}A. Adare {\it et al.}, \Journal{\PRC}{81}{034911}{2010}.
\bibitem{staromega}  L. Adamczyk {\it et al.}, \Journal{\PRL}{113}{022301}{2014}; L. Adamczyk {\it et al.}, a longer version, to be submitted to Phys. Rev. C.
\bibitem{rapp:09}R. Rapp, J. Wambach and H. van Hees, in Relativistic Heavy-Ion Physics, edited by R. Stock and Landolt B\"{o}rnstein (Springer), New Series I/23A (2010) 4-1 [arXiv:0901.3289[hep-ph]].
\bibitem{PSHD:12} O. Linnyk {\it et al.}, \Journal{\PRC}{85}{024910}{2012}.
\bibitem{USTC:12} H. Xu {\it et al.}, \Journal{\PRC}{85}{024906}{2012}.
\bibitem{thermalphoton}A. Adare {\it et al.}, \Journal{\PRL}{104}{132301}{2010}.
\bibitem{Art:98} A.M. Poskanzer and S.A. Voloshin,  \Journal{\PRC}{58}{1671}{1998}.
\bibitem{photonv2} A. Adare {\it et al.}, \Journal{\PRL}{109}{122302}{2012}. 
\bibitem{rapp:11}H. van Hees, C. Gale, R. Rapp, \Journal{\PRC}{84}{054906}{2011}.
\bibitem{Gale:07} R. Chatterjee {\it et al.},
\Journal{\PRC}{75}{054909}{2007}.
\bibitem{phenixcharm:09} A. Adare {\it et al.}, \Journal{\PLB}{670}{313}{2009}.
\bibitem{startof}B. Bonner {\it et al.},
\Journal{\NIMA}{508}{181}{2003}; M. Shao {\it et al.},
\Journal{\NIMA}{492}{344}{2002}; J. Wu {\it et al.},
\Journal{\NIMA}{538}{243}{2005}.
\bibitem{stardaq} J.M. Landgraf {\it et al.}, \Journal{\NIMA}{499}{762}{2003}.
\bibitem{star} K. H. Ackermann {\it et al.}, \Journal{\NIMA}{499}{624}{2003}.
\bibitem{startpc} M. Anderson {\it et al.}, \Journal{\NIMA}{499}{659}{2003}.
\bibitem{bichsel} H. Bichsel, \Journal{\NIMA}{562}{154}{2006}.
\bibitem{pidpp08} Y. Xu {\it et al.}, \Journal{\NIMA}{614}{28}{2010}.
\bibitem{pidNIMA}M. Shao {\it et al.}, \Journal{\NIMA}{558}{419}{2006}.
\bibitem{tofPID} J. Adams {\it
et al.}, \Journal{\PLB}{616}{8}{2005}; L. Ruan, Ph. D. thesis,
USTC, 2005, nucl-ex/0503018.
\bibitem{starvpd} W.J. Llope {\it et al.},  \Journal{\NIMA}{522}{252}{2004}.
\bibitem{centraltag} C. Adler {\it et al.},  \Journal{\PRL}{89}{202301}{2002}.
\bibitem{starelectron} J. Adams {\it et al.},
  \Journal{\PRL}{94}{062301}{2005}.
\bibitem{starppdilepton}  L. Adamczyk {\it et al.}, \Journal{\PRC}{86}{024906}{2012}.

\bibitem{Jie:13} J. Zhao, Ph.D. thesis, Shanghai Institute of Applied Physics, 2013, https://drupal.star.bnl.gov/STAR/theses/phd-32 .
\bibitem{flattenmethod}  S. A. Voloshin, A. M. Poskanzer and R. Snellings, in
Landolt-Boernstein, Relativistic Heavy Ion Physics, Vol.
1/23, p. 5-54 (Springer-Verlag, 2010). arXiv:0809.2949
[nucl-ex].
\bibitem{starv2:13} L. Adamczyk {\it et al.}, \Journal{\PRC}{88}{014902}{2013}. 


\bibitem{ksv2} B.I. Abelev {\it et al.}, \Journal{\PRC}{77}{054901}{2008}.
\bibitem{AuAuPID} B.I. Abelev {\it et al.}, \Journal{\PRL}{97}{152301}{2006}.
\bibitem{pidlowpt} B.I. Abelev {\it et al.},
  \Journal{\PRC}{79}{034909}{2009}.
\bibitem{starphi} B.I. Abelev {\it et al.}, \Journal{\PRC}{79}{064903}{2009}; J. Adams {\it et al.}, \Journal{\PLB}{612}{181}{2005}.
\bibitem{phenixeta} S. S. Adler {\it et al.}, \Journal{\PRC}{75}{024909}{2007}.
\bibitem{Tsallis} Z. Tang {\it et al.},
\Journal{\PRC}{79}{051901}{2009}; M. Shao {\it et al.},
\Journal{\JPG}{37}{085104}{2010}.
\bibitem{piv2} S. Afanasiev {\it et al.}, \Journal{\PRC}{80}{054907}{2009}.
\bibitem{starpiv2} J. Adams {\it et al.}, \Journal{\PRC}{72}{014904}{2005}.
\bibitem{phiv2}  B.I. Abelev {\it et al.}, \Journal{\PRL}{99}{112301}{2007}.
\bibitem{krollwada} N. M. Kroll and W. Wada, \Journal{\PRV}{98}{1355}{1955}.
\bibitem{ruan:11} L. Ruan {\it et al.},
\Journal{\NPA}{855}{269}{2011}; B. Huang, Ph. D. thesis, USTC, 2011.
\bibitem{pythia}T. Sj\"{o}strand {\it et al.}, \Journal{\CPC}{135}{238}{2001}.
\bibitem{Yifei:2012} L. Adamczyk {\it et al.},
\Journal{\PRD}{86}{072013}{2012}.
\bibitem{wei:2011} H. Agakishiev {\it et al.},
\Journal{\PRD}{83}{052006}{2011}.
\bibitem{PHENIXNPE} A. Adare {\it et al.}, \Journal{\PRC}{84}{044905}{2011}.
\bibitem{chargepiv2} A. Adare {\it et al.}, \Journal{\PRC}{85}{064914}{2012}.
\bibitem{hadronv2}  A. Adare {\it et al.}, \Journal{\PRL}{98}{162301}{2007}.
\bibitem{lambdav2}  J. Adams {\it et al.}, \Journal{\PRL}{92}{052302}{2004}.
\bibitem{Gojko:12} G. Vujanovic {\it et al.}, \Journal{\NPA}{904-905}{557c}{2013}; G. Vujanovic {\it et al.}, \Journal{\PRC}{89}{034904}{2014}. 

\end{thebibliography}
\end{document}